\documentclass[10pt,aaspp4]{aastex}

\newcommand\aprmaior{\;\buildrel\hbox{$>$}\over{\sim}\;}    
\newcommand\FRH{Ferraz-Mello et al. 2008}
\newcommand\SFM{Ferraz-Mello 2013}

\slugcomment{To appear in }

\shorttitle{Rotation of stars hosting close-in planets}

\shortauthors{Ferraz-Mello et al.}

\begin{document}



\title{Interplay of tidal evolution and stellar wind braking in the rotation of stars hosting massive close-in planets}







\author{S. Ferraz-Mello, M. Tadeu dos Santos, H. Folonier} 

\affil{Instituto de Astronomia Geof\'isica e Ci\^encias
 Atmosf\'ericas, Universidade de S\~ao Paulo, SP 05508-090, Brasil} 
\email{sylvio@iag.usp.br}

\author{Sz. Csizmadia}

\affil{Institut f\"ur Planetenforschung, Deutsches Zentrum f\"ur Luft und Raumfahrt (DLR), D-12489 Berlin, Germany}
\author {J.-D. do Nascimento Jr}
\affil {Harvard-Smithsonian Center for Astrophysics, Cambridge MA 02138, USA, \\ and Dep. Fisica Te\'orica e Experimental, Universidade Federal do Rio Grande do Norte, Natal 59072-970, Brasil} 
\and 
\author{M.P\"atzold} \affil {Rheinisches Institut f\"ur Umweltforschung, Universit\"at zu K\"oln, 
D-50931 K\"oln, Germany}










\begin{abstract}

This paper deals with the application of the creep tide theory (Ferraz-Mello, Cel. Mech. Dyn. Astron.  \textbf{116}, 109, 2013) to the study of the rotation of stars hosting massive close-in planets. The stars have nearly the same tidal relaxation factors as gaseous planets and the evolution of their rotation is similar to that of close-in hot Jupiters: they tidally evolve towards a stationary solution. However, stellar rotation may also be affected by stellar wind braking. Thus, while the rotation of a quiet host star evolves towards a stationary attractor with a frequency ($1+6e^2$) times the orbital mean motion of the companion, the continuous loss of angular momentum in an active star displaces the stationary solution towards slower values: Active host stars with big close-in companions tend to have rotational periods longer than the orbital periods of their companions. The study of some hypothetical examples shows that, because of tidal evolution, the rules of gyrochronology cannot be used to estimate the age of one system with a large close-in companion, no matter if the star is quiet or active, if the current semi-major axis of the companion is smaller than 0.03--0.04 AU. Details on the evolution of the systems:  CoRoT LRc06E21637, CoRoT-27, Kepler-75, CoRoT-2, CoRoT-18, CoRoT-14 and on hypothetical systems with planets of mass 1--4 M$_{\rm Jup}$ in orbit around a star similar to the Sun are given.
\end{abstract}







\section{Introduction}

Bright stars with transiting planets unraveled by the space telescopes CoRoT (Barge et al. 2008) and Kepler (Borucki et al. 2010) are objects of choice for the study of characteristic evolutionary parameters because of:

\begin{itemize}
\item The better signal-to-noise ratio of space measurements;
\item Continuous photometry for long time spans allowing the determination of a photometric period not affected by the inclination indetermination of the spectroscopically determined $V \sin I$, which may inform about the rotation of the star notwithstanding the fact that some unexpected commensurabilities may be showing that the rotation of the photometric features is perturbed by the existence of the close-in transiting planet (see B\'eky et al. 2014);
\item The possibility of follow-up observations using high-quality spectrographs. The combination of velocimetry and transit data allows the determination of the true planetary mass and radius (and density). Besides, the joint analysis of both data sets allows a better determination of the orbital eccentricity which, when only velocimetry data are available, shall be determined only from the asymmetries of the radial velocity curve.
\end{itemize}

The study of the past evolution of extrasolar systems with close-in planets in well-determined elliptic orbits is instrumental in constraining the values of the dissipation in the planet, which is an important factor in the tidal evolution. 
However, {in evolution studies constrained only by the current state of the system and a tidal evolution model, often the past rotation of the star becomes too slow in contradiction with the accepted models for the primordial rotation.} 
In addition, evolved systems in which the current star's rotation is much slower than the period of the companion cannot be explained by considering only the tidal evolution. It is clear that some braking process may be acting at the same time as the tidal torque and that the evolution of the stellar rotation may be determined by the interplay of both processes.

In this paper, we study the evolution of the rotation of a star under the joint effects of the tides raised on it by a close-in hot Jupiter or brown dwarf and the stellar wind braking. 

There is observational evidence of tidal evolution in stars hosting massive close-in planets  (Pont, 2009, Lanza, 2010). 
In the absence of braking, the star's rotation evolves towards a quasi-synchronous attractor. 
This is the classical Darwinian result (Hut, 1981; Ferraz-Mello et al. 2008; Williams and Efroimsky, 2012). However, the star's continuous loss of angular momentum due to the stellar wind displaces the stationary solution towards sub-synchronous rotation values. As a consequence, active solar-type host stars with big close-in companions tend to have rotational periods longer than the orbital periods of their companions, but smaller than the values predicted by the gyrochronology (Barnes, 2007) for stars without close-in companions.  
On the other hand, F stars with larger masses follow the Darwinian solution and have nearly synchronous rotations. 

The tidal evolution theory adopted in the present study is the creep tide theory (Section 2). 
However, the choice of a given tidal evolution theory does not affect the results. 
Each tidal evolution theory introduces an undetermined parameter that may be freely adjusted.
It is the relaxation factor of the creep tide theory (\SFM) used here, the tidal lag of Darwin's theory (see \FRH), the dissipation factor $1/Q$ (see P\"atzold 2012), Mignard's time-lag (Mignard 1979) or even the dissipation parameter introduced by Eggleton et al (1998). 
However, the main results of these theories are functionally similar when the considered bodies are stars or giant gaseous planets, and the unknown parameters can be adjusted so that their results become similar.
{The creep tide theory has been selected because it depends on only one free parameter (the relaxation factor $\gamma$) and its expansions can be extended to the order of approximation necessary to obtain results that are numerically precise, even when high eccentricities are involved.}

In stars hosting planetary systems with massive close-in companions, such as hot Jupiters
or brown dwarfs, the tidal evolution of the system cannot be studied without taking into
account a possible braking of the star's rotation, which affects the location of the stationary
solutions, displacing them toward subsynchronous values.
The magnetic stellar wind braking is modeled here (Section 3) with the semi-empirical formulas derived by Bouvier and collaborators (Bouvier et al., 1997; Bouvier, 2009, 2013), which reproduces well the great uniformity of the braking process and its limits, clearly seen in the plots of photometric rotation period versus color indices of the open  {clusters NGC 6811 and NGC 6819} based on observations made by the space telescope Kepler (Meibom et al. 2011, 2015). 

In the second half of the paper, we present results on the evolution of the rotation of several stars with large transiting companions under the action of both tides and braking. 
We put a major emphasis on CoRoT LRc06E21637 (Sec. \ref{Sec:C3x}). 
This star was chosen as a paradigm because its companion is a big brown dwarf and its estimated age is large enough to allow the study of the process of convergence of the star's rotation to a stationary value.
We also present in Sec. \ref{Sec:C27K75} and Sec. \ref{Sec:C2C18} the results on some other G stars with emphasis on CoRoT-27, Kepler-75, CoRoT-2 and CoRoT-18. In Section \ref{Sec:F-stars}, we discuss the rotation of three F stars: CoRoT-15, KELT-1  and HAT-P-2. The stationary value reached by the rotation of these stars is determined solely by the tidal friction (the braking is absent). The study of the F9 star CoRoT-14 indicates the presence of braking, but gives rise to some unexpected relaxation values. A shorter rotation period, however, would allow us to put this star in the same class as CoRoT-2 and CoRoT-18, but with a larger lifetime (Sec. \ref{Sec:C14}). The basic parameters of the studied stars and their companions are presented in Table \ref{table}. In general, the values shown are taken from the cited references but, in a few cases, as indicated, they were redetermined from the photometric and velocimetric sets of observational data. 

Finally, Sec. \ref{Sec:geral} considers hypothetical systems with close-in companions of mass 1--4 $M_{\rm Jup}$ and sets the distance limits for the existence of significant tidal torques. The results shown correspond to an initially fast rotating star with the typical period 2 days and one companion in circular orbit. The extensions of the results to slower stars and/or with companions in non-circular orbits are also discussed.

\section{Tidal evolution}
In the creep tide theory the rotation of a central body perturbed by tidal torques of an external body in Keplerian motion in the plane of its equator is such that
\begin{equation}\label{eq:omegadot}
\frac{d\Omega}{dt} =  \frac{3GM}{2r^{3}}\overline\epsilon_\rho 
\sum_k E_{2,k} \cos\overline\sigma_k \sin\big(2\varphi+(k-2)\ell-2\omega-\overline\sigma_k\big)
\end{equation}
where $G$ is the gravitational constant, $M$ is the mass of the external body, $r$ is the distance between the two bodies, $\ell$ and $\varphi$ are the mean and true anomalies, $\omega$ is the argument of the pericenter,
\begin{equation}
\overline\epsilon_\rho=\frac{15}{4}\Bigg{(}\frac{M}{m}\Bigg{)}\Bigg{(}\frac{R}{a}\Bigg{)}^3
\end{equation}
is the prolateness induced by the tides in a homogeneous\footnote{N.B. The torque and the equations of the rotation are, however, calculated using the actual moment of inertia of the star.} body of mass $m$ ($a$ is the semi-major axis), and 
\begin{equation}\label{eq:Cayley}
E_{q,p}(e)=\frac{1}{2\pi}\int_0^{2\pi}\left(\frac{a}{r}\right)^3
\cos\big(qv+(p-q)\ell\big)\ d\ell.
\end{equation}
are the so-called Cayley functions (see Ferraz-Mello, 2013).
Terms of second order with respect to $\overline\epsilon_\rho$ are neglected in the calculations. 

The $\overline\sigma_k$ are phase angles related to the integration of the first-order differential equation of the creep, 
\begin{equation}
\overline\sigma_k=\arctan \left(\frac{\nu+kn}{\gamma}\right)
\end{equation}
\begin{equation}
\sin \overline\sigma_k = \frac{\nu+kn}{\sqrt{\gamma^2+ (\nu+kn)^2}},
\hspace{1cm}
\cos \overline\sigma_k = \frac{\gamma }{\sqrt{\gamma^2+ (\nu+kn\big)^2}}
\end{equation}
where 
\begin{equation}
\nu=2\Omega-2n
\end{equation} 
is the semi-diurnal frequency and $\gamma$ is the creep relaxation factor
related to the physical parameters of the central body through
\begin{equation}\label{eq:eta}
\gamma=\frac{w R}{2\eta} = \frac{3gm}{8\pi R^2 \eta},
\end{equation}
where $w$ and $g$ are the specific weight and gravity acceleration on the surface of the body, $\eta$ the uniform viscosity, and $R$ is its mean radius.
When $\gamma \gg \nu$, the tangent of the angle $\overline\sigma_0$ is proportional to the semi-diurnal frequency and thus, in this case, $\overline\sigma_0$ is akin to the ad hoc phase lag introduced in classical versions of Darwin's theory.

In the context of this paper, the central body is the star and the orbiting external body is its companion exoplanet or brown dwarf. However, Eqn. (\ref{eq:omegadot}) is also valid for the rotation of the planet if the roles played by the two bodies are inverted. 

One important characteristic of this equation is that the right-hand side is independent of the position of the body. 
The arguments of the periodic terms are independent of the body rotation angle. It is a true first-order differential equation and, unlike the rigid-body spin-orbit approach, the solution does not present free oscillations. 

\subsection{Synchronization. High-$\gamma$ approximation}\label{sec:spinorb}

Eqn. (\ref{eq:omegadot}) is the first-order differential equation giving, in the creep tide theory, the variations of the rotation of the star due to the tide raised by the close-in companion\footnote{We do not use the words `resonance' and `capture' because the dynamics of this approach is not pendulum-like. We rather have attractors and basins of attraction. }. 
For a generic order $N$, the Keplerian approximation may be written as:
\begin{equation}
\frac{d\Omega}{dt} = \ -\frac{3GM\overline\epsilon_\rho }{4a^{3}}\sum_k 
E_{2,k}   \sum_j 
E_{2,k+j} \left(
\sin 2\overline\sigma_k \cos j\ell + 2\cos^2\overline\sigma_k \sin j\ell\right).
\label{eq:yprime}
\end{equation}
where the summation limits are to be fixed taking into account that 
$E_{2,k}\sim {\cal{O}}(e^{|k|})$ and be such that all terms of orders lower than a prefixed order $N$ are included. 

If $\gamma \gg n$ (as in the case of giant planets and stars) and $\Omega = {\cal O}(n)$, then $\nu+kn \ll \gamma $ and  we may use the high-$\gamma$ approximations
\begin{displaymath}
\cos \overline\sigma_k \approx 1,
\qquad
\sin \overline\sigma_k = \frac{\nu+kn}{\gamma}
\end{displaymath}
and
\begin{equation}
\frac{d\Omega}{dt} = \ -\frac{3GM\overline\epsilon_\rho }{2a^{3}}\sum_k
E_{2,k}   \sum_j
E_{2,k+j} \left(\frac{\nu+kn}{\gamma}
\cos j\ell +  \sin j\ell\right).
\end{equation}
If we neglect the terms with $j\ne 0$ (short-period terms), the equation is reduced to
\begin{equation}
\frac{d\Omega}{dt} = \ -\frac{3GM\overline\epsilon_\rho}{2a^{3}}
\sum_{k=-N/2}^{N/2} \frac{\nu+kn}{\gamma} E_{2,k}^2   
\label{eq:aver}\end{equation}

Eqn. (\ref{eq:aver}) is easy to solve, giving
\begin{equation}
\nu=\nu_{\rm lim} + \nu_0 e^{-\kappa (t-t_0) }
\end{equation}
where $\kappa$ is the damping coefficient
\begin{equation}
\kappa = \frac{3GM\overline\epsilon_\rho}{\gamma a^{3}}\sum_{k=-N/2}^{N/2} E_{2,k}^2,
\end{equation}
$\nu_0$ is an integration constant and
\begin{equation}
\nu_{\rm lim}=-\frac{n \sum_k kE_{2,k}^2}{\sum_k E_{2,k}^2}=12 ne^2 + \frac {3}{4} ne^4 + \frac{173}{4}ne^6 + \cdots.
\end{equation}
Therefore, the rotation velocity tends to 
\begin{equation}
\Omega_{\rm lim}=n\Big(1+6e^2 + \frac{3}{8} e^4 + \frac{173}{8}e^6 + \cdots \Big),
\label{eq:offset}\end{equation}
which is the same limit given by the classical theories {following Darwin's basic assumptions} (see e.g. Hut, 1981, Eqn. 42; Williams and Efroimsky, 2012, Eqn. 44)\footnote{There is a  misprint in Hut's Eqn. 43 where the coefficient of $e^6$ appears as 223/8. 
The expansion of his Eqn. 42 gives, for this coefficient, the value 173/8, as here.}.

These  stationary rotations are often called ``pseudosynchronous" or ``supersynchronous". Supersynchronous  means here that the rotational angular velocity is larger than the mean motion, and the actual orbit of the companion lies outside the so-called synchronous orbit -- the one with same mean motion as the stellar rotation velocity. 

{During the time span of the studied star's rotations, the semi-major axis and the eccentricity of the companion vary. To take this variation into account, we have also used the averaged equations of the creep tide theory:}
\begin{equation}
\frac{da}{dt}=\frac{3 k_f  M R^5  n}{4m a^4}
\sum_{k} (2-k)E_{2,k}^2(e)\sin 2\sigma_k.
\end{equation}
\begin{equation}\label{eq:doteav}
\frac{de}{dt}=-\frac{3k_f MR^5n}{8ma^5e}\sum_{k} \Big(2\sqrt{1 - e^2} - (2-k)(1 - e^2)\Big)E_{2,k}^2(e)\sin 2\sigma_k.
\end{equation}
{These equations give the variation of the elements due to the tides raised in the star. 
Two other identical equations with the meanings of the two bodies inverted were also added to the codes to give the variations due to the tides raised in the companion. However, the contribution of the tides in the companion is negligible, in the studied cases, because of the quick trapping of the companion's rotation to a stationary pseudo-synchronous state. }

\subsection {Short-period oscillations. }

\begin{figure}[t]
\centerline{\hbox{
\includegraphics[width=6cm,clip=]{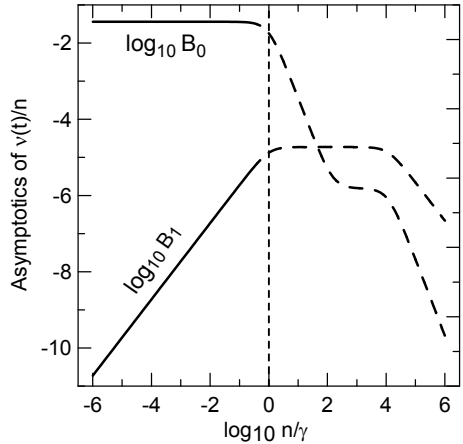}
}}
\caption{Mean ($B_0$) and amplitude ($B_1$) of the asymptotic periodic solution of Eq. (\ref{eq:yprime}) as a function of $\gamma$. The solid lines correspond to $\gamma>n$, i.e. to gaseous planets and host stars. The dashed lines correspond to rocky bodies such as low-mass Earth-like  planets.} \label{fig:B0B1}
\end{figure}

When the terms with $j\ne 0$ of Eqn. \ref{eq:yprime} are considered, the solution becomes more complex, and periodic oscillations whose frequencies are multiple of the orbital period of the system are added to the limit solution. 
Once the transient given by the exponential term is damped, the solution becomes a periodic function. 
We may use the classic technique of undetermined coefficients to construct these periodic terms. 
However, we are only interested in the leading terms of the Fourier expansion. 
Thus, we just assume that $\nu/n=B_0+B_1\cos(nt + {\rm phase})$, substitute it in the differential equation, discard the second- and higher-order harmonics, and solve to obtain $B_0$ and $B_1$. The results are shown in fig. \ref{fig:B0B1} 

On the left-hand side of the plot ($\gamma > n$, i.e $\log_{10} n/\gamma < 0$),  we have the so-called Darwin regime: the asymptotic motion is almost constant and significantly supersynchronous ($\Omega>n$) for non-circular orbits, as already found with the high-$\gamma$ approximation of the previous section. 
In this regime, the offset of the stationary motion ($B_0$) is almost independent of the relaxation factor $\gamma$, and the short-period oscillation ($B_1$) is very small and can be neglected when dealing with stars and giant gaseous planets.

The right-hand side of the plot ($\gamma < n$, i.e $\log_{10} n/\gamma > 0$) corresponds to the tidal regime followed by rocky bodies such as low-mass Earth-like planets. In this case, $\nu$ tends to be very small, that is, the planetary rotation tends to be almost synchronous to the orbital motion notwithstanding the high eccentricity used in the calculation ($e=0.5$). The periodic oscillation ($B_1$) becomes the most important part of the solution 

\section{Braking of the stellar rotation}

{The magnetic stellar wind braking is modeled here with the semi-empirical formulas derived by Bouvier and collaborators (Bouvier et al., 1997; Bouvier, 2009, 2013). 
Even if the braking may have variations throughout the life of the star, mainly in the early phases where the star is still contracting (Penev et al. 2012), the formulas of Bouvier et al. (1997) reproduce fairly well the average evolution of their rotations.
According to these authors, a cubic law is necessary to recover the Skumanich's (1972) relationship,
 $\Omega(t) \propto t^{-1/2}$, which provides a reasonable approximation to the observed braking of slow rotators on the main sequence.
The model differs from the linear model adopted by Dobbs-Dixon et al (2004) in their study of stellar spins in short-period planetary systems.
In Bouvier's model, the linear law is only adopted for rotations faster than a saturation limit. }

Bouvier's model for the stellar braking of low-mass stars ($0.5 M_\odot<m<1.1 M_\odot$), with an outer convective zone, is given by
\begin{equation}
\frac{d\Omega}{dt} = \left\{
\begin{array}{ll}
- f_P B_W\Omega^3&{\rm when}\qquad \Omega\le\omega_{\rm sat}\vspace{3mm}\\
- f_P B_W\omega_{\rm sat}^2\Omega\hspace{1cm}& {\rm when}\qquad \Omega>\omega_{\rm sat}\label{eq:Bouvier}
\end{array}
\right.
\ \end{equation}
where $B_W$ is a factor depending on the star's mass and radius through the relation
\begin{equation}
B_W=2.7\times 10^{47} \frac{1}{C}\sqrt{\Big(\frac{R}{R_\odot} \frac{M_\odot}{m} \Big)}\qquad \qquad   ({\rm cgs \enspace units}).
\label{eq:Bw}
\end{equation}
and $\omega_{\rm sat}$ is the value at which the angular momentum loss saturates {(Bouvier et al. 1997)}, fixed at $\omega_{\rm sat}  = 3, 8, 14 \Omega_\odot$ for 0.5, 0.8, and 1.0 $M_\odot$ stars, respectively. 
The factor $f_P$ was introduced by P\"atzold et al. (2012) in the study of the planet of the subgiant star CoRoT-21 to take into account that the fiducial wind braking given by Bouvier's law may be excessive in that case. 
$C$ is the moment of inertia of the star {(taken as $0.06mR^2$)}. Notwithstanding the fact that the processes which result in the braking take place in the convective zone of the star, some other processes may exist that allow the transfer of angular momentum inside the star, from the inner radiative zone and core to the outer convective zone, so that the whole star rotates with angular velocities of the same order of magnitude. For instance, in the case of the Sun, helioseismological observations reveal that the radiative zone rotates with the same angular velocity as the convective zone at mid latitudes (Garcia et al., 2007).

In the case of the Sun, the coefficient used in Eqn. \ref{eq:Bouvier} corresponds to $6.6  \times 10^{30}$ g cm$^2$ s$^{-2}$. This value may be compared to the estimates of the loss of angular momentum of the Sun which are in the interval $ 4 - 9  \times 10^{30}$ g cm$^2$ s$^{-2}$ (cf. Gallet and Bouvier 2013; but one of the estimates cited by Gallet and Bouvier is $41  \times 10^{30}$ g cm$^2$ s$^{-2}$) while, 
for low-mass M stars, the coefficient in Eqn. \ref{eq:Bouvier} must be taken in the range $1.2 \times 10^{45} - 1.1 \times 10^{47}$ g cm$^2$ s (see Irwin et al. 2011). 

We may also note that the magnetic connection to very close planets may inhibit the stellar wind (Strugarek et al. 2014) and thus reduce the braking torque on the star to a minimum of $f_P=0.7$ for $a\simeq 3R$. This reduction was not used in the study of the solar-type stars considered in this paper since, in all studied cases, we have $a \aprmaior 5R$. 
The other torques due to the star-planet magnetic interaction (see Strugarek et al.2014) were also neglected. In the case of closest planets, the non-consideration of these smaller torques may increase the uncertainty in the values estimated for the relaxation factor $\gamma$.


For the application concerned here, it is sufficient to know that the above form of the law is valid after the star has completed its contraction (the stellar moment of inertia $C$ no longer changes significantly), is fully decoupled from the disk, and no significant mass loss needs to be considered. 
The adopted braking model {relies upon a simplified approach and adopts simple power laws for how the magnetic field strength and the wind velocity vary with distance from the star, which is at variance with the more complete model recently proposed (Matt et al. 2012, 2015) that assumes a bi-dimensional dipolar magnetic geometry at the stellar surface. However, the new model depends on some additional ill-known parameters such as the mass loss rate and the stellar magnetic field, as well as two free adimensional parameters. The choice of Bouvier's formulas for the present study is justified by its successful applications } in the study of the evolution of the rotation of single stars (see Bouvier 2009, 2013), which allows us to expect that it gives correct values also in the case where the star is also in tidal interaction with a close-in companion.


\subsection{The stationary solution}

\begin{figure}[t]
\centerline{\hbox{
\includegraphics[height=6cm,clip=]{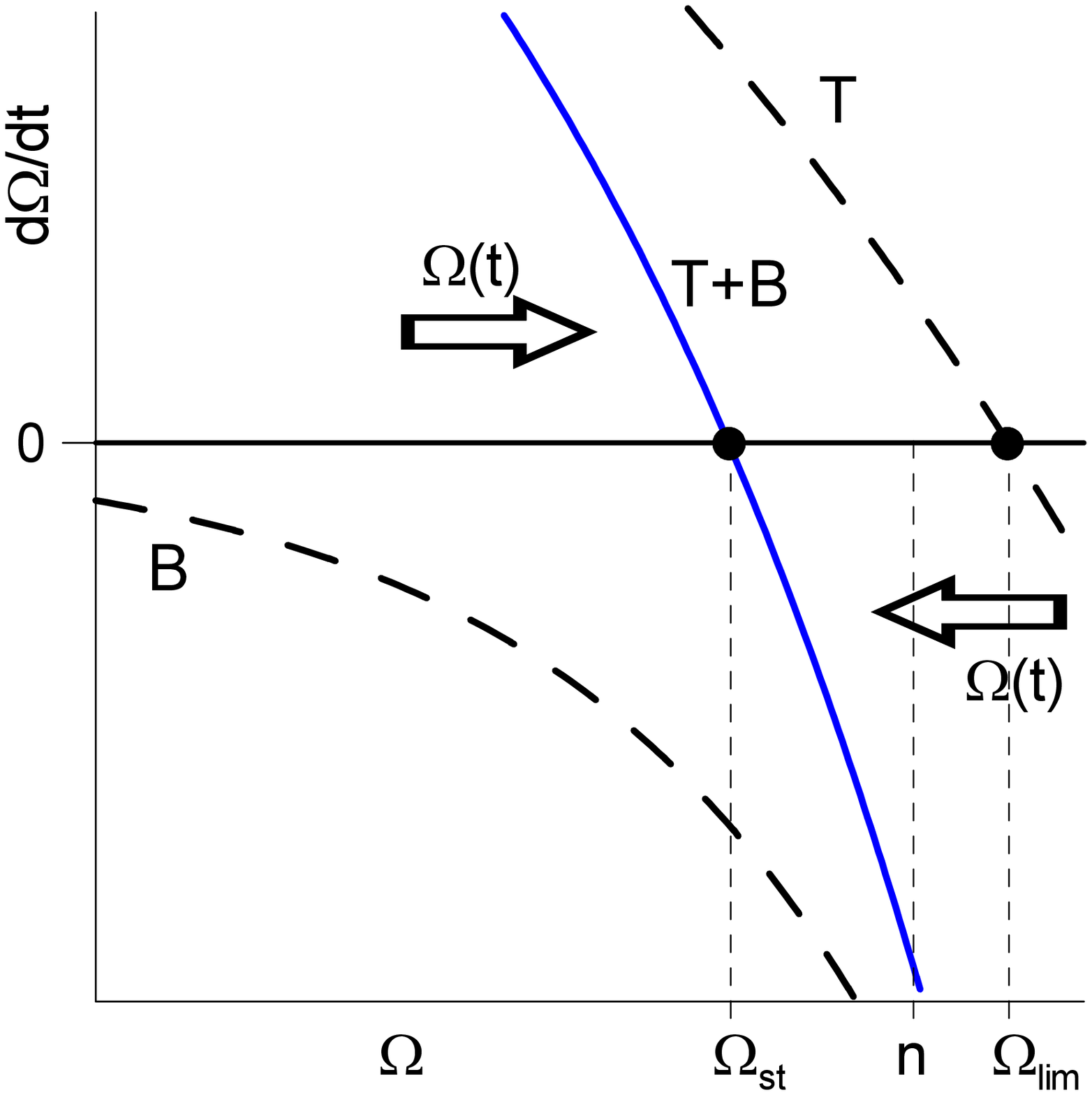}\hspace{20mm}
\includegraphics[height=6cm,clip=]{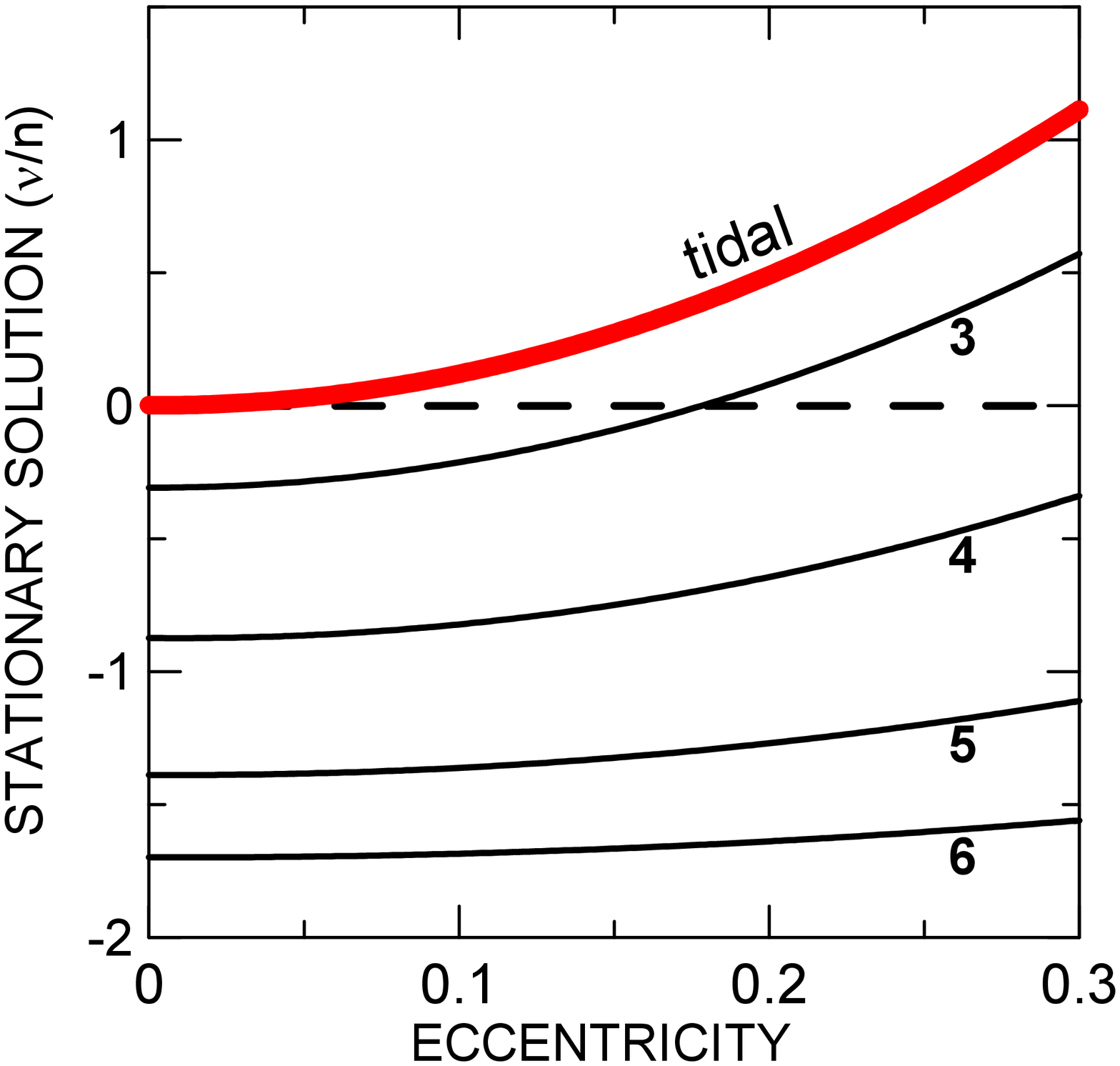}
}}\caption{\textit{Left}: {Solid line: Plot of the function $\dot{\Omega}(\Omega)$ resulting from the joint action of the tidal torque (T) and the magnetic braking (B). The intersection with the axis $\dot{\Omega}=0$ is the stable stationary solution $\Omega_{\rm st}$. 
$\Omega=\Omega_{\rm lim}$ is the solution in the absence of braking.}
\textit{Right}: Location of the corresponding stable stationary solutions $\nu_{\rm st}$ and $\nu_{\rm lim}$ as functions of the orbital eccentricity, for several values of the relaxation factor.
The stationary solutions were calculated adopting $\log \gamma/n$ = 3, 4, 5, 6 respectively (see labels).
{N.B. $\nu=-2n$ is equivalent to $\Omega=0$.}}
\label{fig:stat}
\end{figure}

{One solution in which the total torque acting on the star vanishes is called stationary.
Figure \ref{fig:stat} (\textit{left}) shows the functions $\dot{\Omega}(\Omega)$ defined by Eqs. (\ref{eq:aver}) and (\ref{eq:Bouvier}) (for $\Omega\le\omega_{\rm sat}$) (dashed lines) and their sum (solid line). 
In the upper(resp. lower) half-plane, the torque is positive (resp.negative) and the angular velocity of the star rotation is increasing (resp. decreasing). When the braking is not present, the equation $\dot\Omega = 0$ has a stable solution at $\Omega=\Omega_{\rm lim}\ (\ge n)$. The corresponding motion is synchronous if $e=0$ and supersynchronous otherwise (see Eq. \ref{eq:offset}). When the braking is considered, the stable root is displaced to $\Omega_{\rm st} \  (< \Omega_{\rm lim})$. }

Fig. \ref{fig:stat} (\textit{right}) shows the location of the corresponding stable stationary solutions $\nu_{\rm st}=2(\Omega_{\rm st}-n)$ and $\nu_{\rm lim}=2(\Omega_{\rm lim}-n)$ as functions of the orbital eccentricity, for several values of the relaxation factor. The used values of $\log \gamma/n$ label the curves shown.
{The stable solutions $\Omega_{\rm st}$ are generally subsynchronous (i.e $\Omega_{\rm st}<n$), but they can be supersynchronous if the eccentricity and the relaxation factor are high.}

\begin{figure}[t]
\centerline{\hbox{
\includegraphics[width=6cm,clip=]{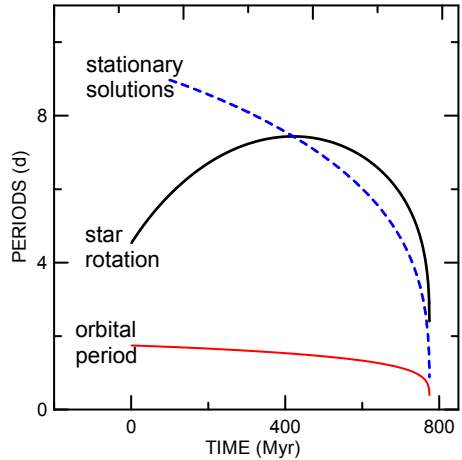}
}}
\caption{ {Evolution of CoRoT-2's rotation period (black), of the orbital period of its planet (red), and of the stationary solution (dashed blue). $\gamma$ = 100 s$^{-1}$. } }\label{fig:PerStat}
\end{figure}

It is worth emphasizing that stationary solutions are not static; they evolve as the parameters of the system evolve. Subsynchronous stationary solutions cause the tidal decrease of the semi-major axis and the consequent migration of the stationary solution itself towards smaller periods. {When a massive companion is too close to the star, the migration of the stationary solution may be so fast that the system reaches the stationary solution, but is not trapped by it. This is the case of CoRoT-2 and CoRoT-18 (see Section \ref {Sec:C2C18}). Figure \ref{fig:PerStat} shows the evolution of CoRoT-2 when $\gamma$= 100 s$^{-1}$. It shows the star's rotation being driven by the stationary solution. In the first half the rotation period  increases up to reach the stationary solution at around $t \sim$ 400 Myr, crosses it, and then decreases evolving again towards the downward migrating stationary solution, up to the eventual fall of the planet into the star.}

\begin{deluxetable}{lcccccclc}
\tabletypesize{\scriptsize}

\rotate

\tablecaption{Basic parameters of the systems discussed in this paper}\label{table}
\tablewidth{0pt}

\tablehead{
\colhead{Star} & \colhead{Spectral} & \colhead{Star's mass } & \colhead{Companion's} & \colhead{Orbital period} & \colhead{Rotation period} &
\colhead{Adopted $\gamma$} & \colhead{Age } &\colhead{{Eccentricity}}\\

\colhead{} &\colhead{type} & \colhead{($M_\odot$)} & \colhead{mass ($M_{\rm Jup})$} & \colhead{(d)} & \colhead{ (d)} &
\colhead{(s$^{-1}$)} & \colhead{ (Gyr)} \\}
\startdata

LRc06E21637$^a$&G9V& $0.93^{+0.05}_{-0.02}$ &  $62\pm 5$ & 5.81893(63)  & $8.95\pm 0.04 \dag^m$ & 55 & 4.6$-$11&0.070(3) \\
Kepler-75$^b$ &  G8V & $0.88 \pm 0.06 $& 9.9 $\pm$ 0.5 & 8.884924(2) &  19.2 $\pm$ 0.3 $\dag$ & 108  & 6 $\pm$ 3 & 0.57(10) \\
CoRoT-27$^c$ &   G2 & $1.05 \pm 0.11 $&  10.39 $\pm$ 0.55&  3.57532(6) & 12.7 $\pm1.7 \ddag$    & 90  &  4.2 $\pm$ 2.7 & $<$ 0.065\\
HAT-P-20$^d$ &   K3 & $0.756\pm 0.028$&  7.246 $\pm$ 0.187&  2.875317(4) & 16.7 $\pm 4.5 \ddag$ & 40  &  $6.7^{+5.7}_{-3.8}$ & 0.0150(5)\\
HAT-P-21$^d$ &   G3 & $0.947 \pm 0.042$&	  4.063 $\pm$ 0.161&  4.124461(7) & 13.63 $\pm 3.1 \ddag$ & 25  &  $10.2 \pm 2.5$ & 0.228(16)\\
CoRoT-2$^e$ & G7V & $ 0.97 \pm 0.06$&3.31 $\pm$ 0.16 &  1.742996(2) & 4.52 $\pm$ 0.024$\dag$ & 20$-$100  &  0.2 $-$ 4.0 & $<$ 0.06\\
CoRoT-18$^{f,g}$& G9V & $0.95 \pm 0.15$ & 3.47 $\pm$ 0.38 &  1.900070(3) & 5.4 $\pm$ 0.6 $\dag$ & 20$-$100 &  0.05$-$1.0& $<$ 0.08 \\
CoRoT-15$^h$& F7V  &1.32$ \pm$ 0.12 & 63.3 $\pm$ 4.1 &  3.06036(3) & $3.9^{+1.2}_{-0.8}   \ddag\ \S$ &&  1.14$-$3.35 & 0(?) \\
KELT-1$^i$&   F5	& $1.335 \pm 0.063$ & 27.38 $\pm$ 0.93 &  1.217514(15)& 1.35 $\pm 0.04  \ddag$  &&  1.5 $-$ 2 & 0(?)\\			
HAT-P-2$^{j,k}$&   F8& $1.36 \pm 0.04$ & 8.74 $\pm$ 0.26 &  5.633473(6) & 4.4 $\pm 0.7  \ddag$  &&  2.6 $\pm$ 0.7 & 0.517(3)\\			
CoRoT-14$^{l}$& F9V & $1.13\pm 0.09$  &6.94 $\pm$ 0.5$^m$ &  1.51214(13) & $5.6 \pm 0.5 \dag^m$ & &  0.4 $-$ 8.0 & 0(?) \\
KOI-205$^{n}$& K0 & $0.925\pm 0.033$& $39.9\pm1.0$&  11.720125(2) & $40.99 \pm 0.5 \dag$ &   $ > 100$ & 0.4--8.3 & $<$  0.03 \\
\hline\\
\multicolumn{7}{l}{$\dag$ photometric; $\ddag$ $\times \sin I$;
$\S$ photometric modulations at 2.9, 3.1 and 6.3 days}\\

\\
\enddata


%
\tablerefs{
(a) Sz. Csizmadia et al., in preparation (Proposed name: CoRoT 33); (b) H\' ebrard et al. (2013); (c) Parviainen et al. (2014); (d) Bakos et al. (2011); (e) Alonso et al. (2008); (f) H\' ebrard et al. (2011); (g) Moutou et al. (2013); (h) Bouchy et al. (2011); (i) Siverd et al (2012); (j) Southworth (2010); (k) Albrecht et al. (2012); (l) Tingley et al. (2011);  (m) this paper; (n) D\'{\i}az et al. (2013). }
\end{deluxetable}

\section{The CoRoT candidate LRc06E21637 and its brown dwarf companion. A paradigm}\label{Sec:C3x}
The best example of a system whose evolution is determined by both a strong tide and the stellar wind braking is the CoRoT candidate LRc06E21637 where the companion is a massive brown dwarf and the tidal effects are enhanced. The preliminary parameters of the system were determined by the CoRoT exoplanets science team (Cest) and the main ones are shown in Table 1. It is worth mentioning that the photometric period is 3/2 of the orbital period of the companion, reinforcing the idea that the rotation of the features determining the photometric periods is in some unknown way modulated by the orbital motion of the companion (see B\'eky et al, 2014). 

\begin{figure}[t]
\centerline{\hbox{
\includegraphics[height=4.5cm,clip=]{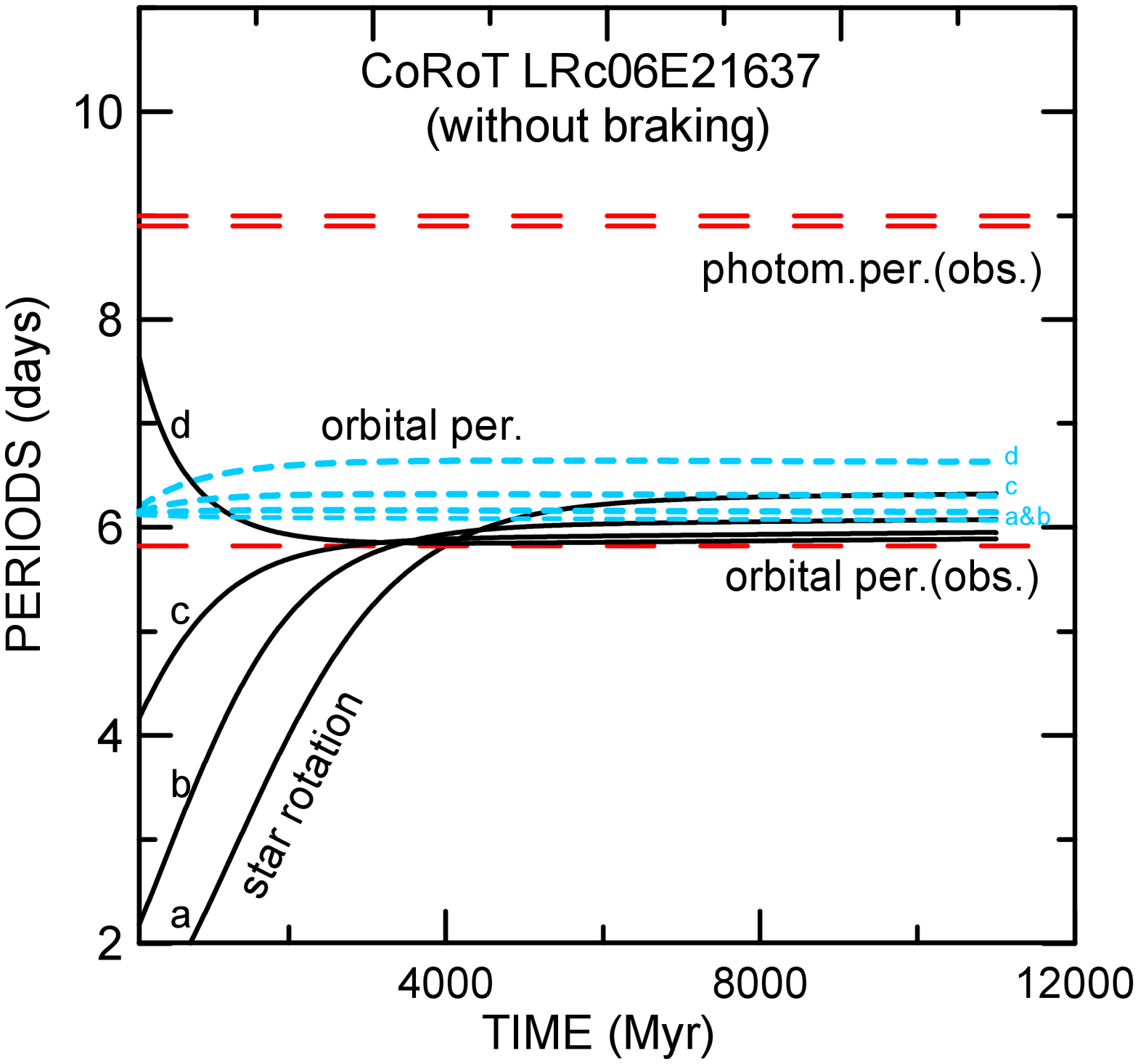}\hspace{3mm}
\includegraphics[height=4.5cm,clip=]{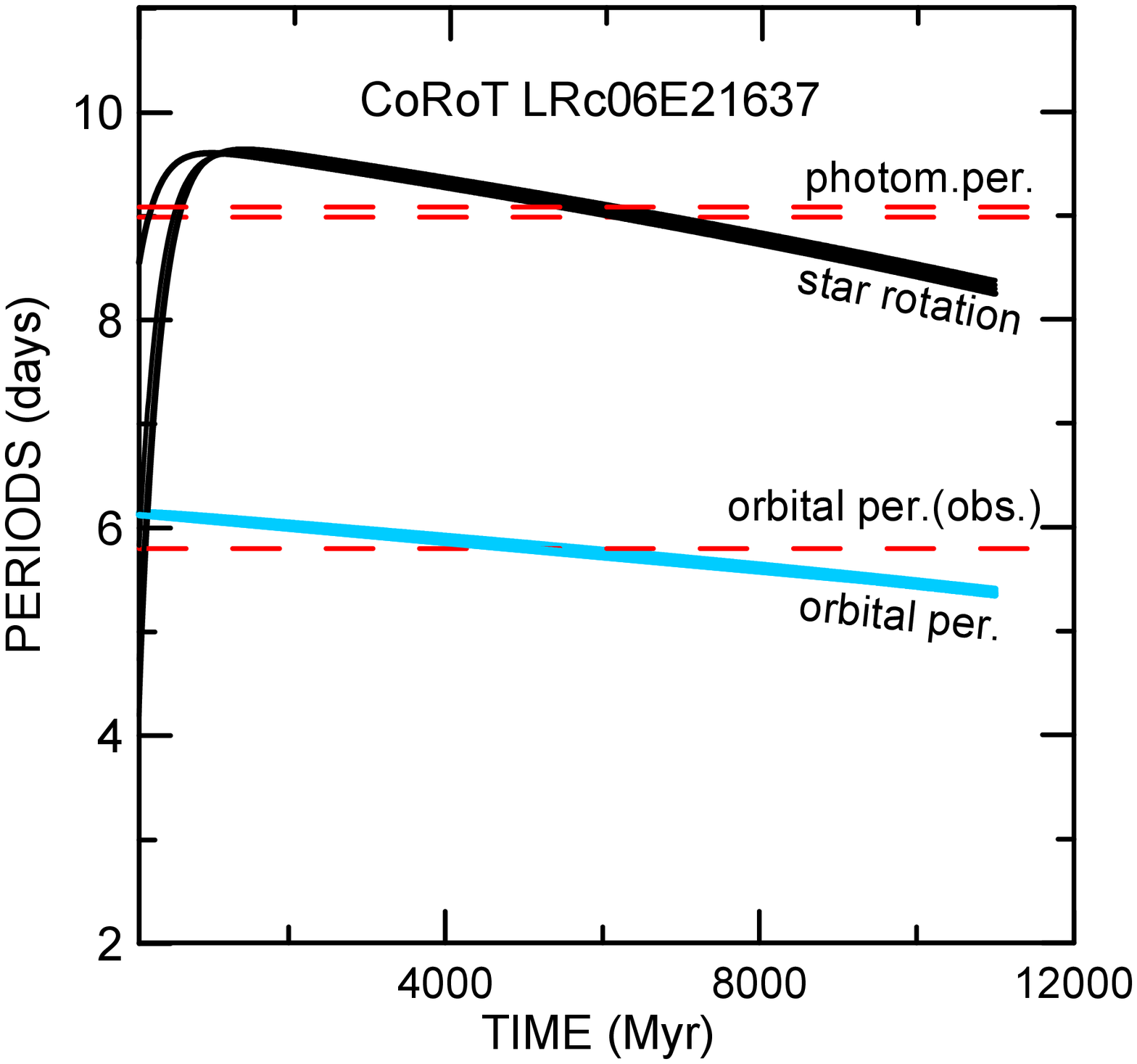}\hspace{3mm}
\includegraphics[height=4.5cm,clip=]{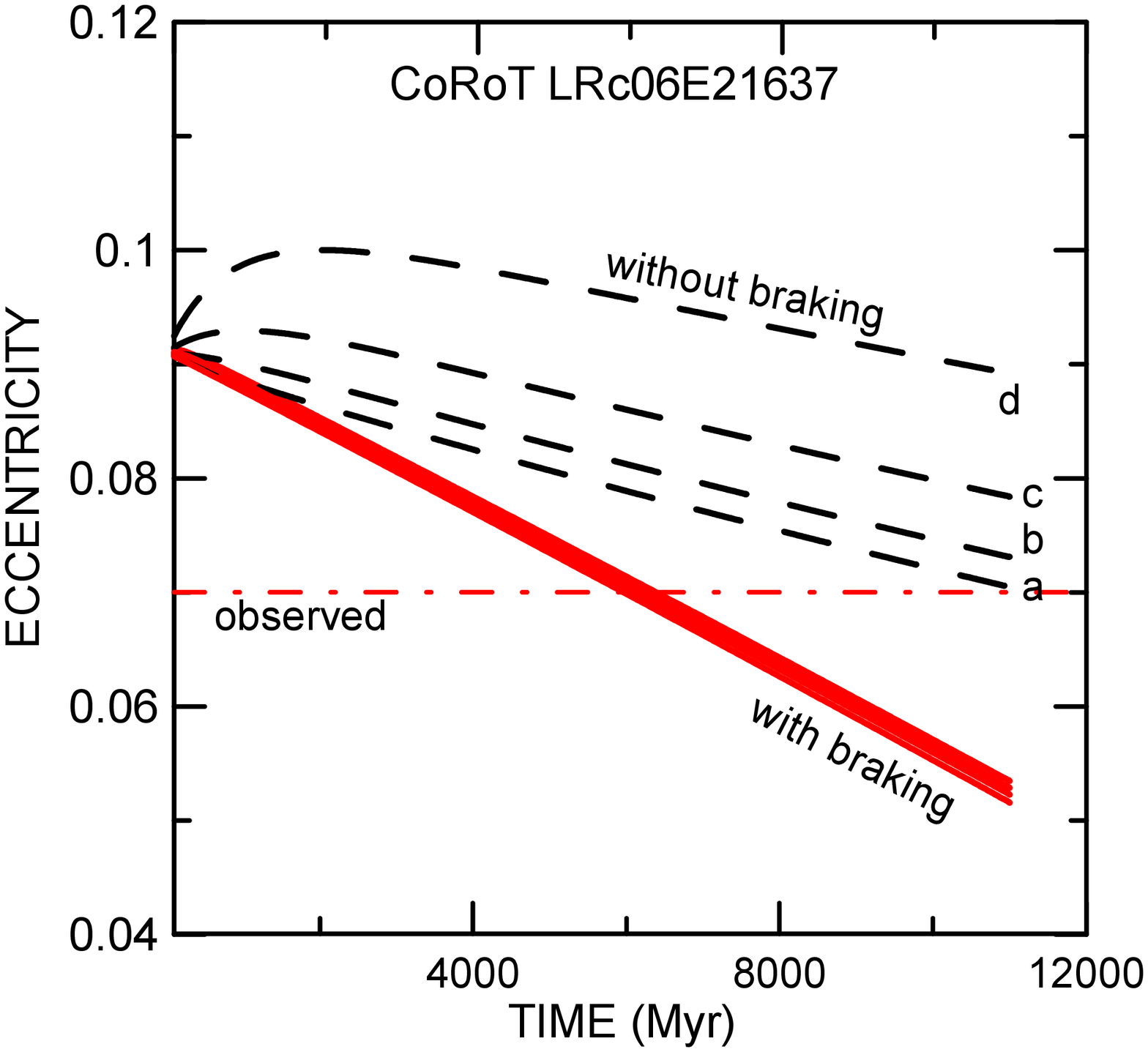}
}}
\caption{{\it Left}: Evolution of the star's rotation and of the orbital period of the companion in the classical approach when the stellar rotation is only affected by tides. {\it Middle:} Evolution resulting from the joint action of the tidal torque on the star and the stellar wind braking.   The horizontal dashed lines show the observed values of the orbital period of the companion and the photometric period of the star. {\it Right:} Evolution of the eccentricity in the two cases. Four different values of the initial period of rotation of the star were considered: 1, 2, 4, and 8 days,  {labeled, $a, b, c, d$, respectively.} }
\label{fig:Corot3x}
\end{figure}

The system is a paradigm of the interplay of tidal evolution and stellar wind braking because of the high mass of the companion, the distance of the companion from the star ($a=0.0623$ AU), allowing it to evolve {notably}, and an age allowing the star to have already reached the subsynchronous stationary configuration.  The present variation of the stellar rotation is  due to the fact that the orbital period of the companion is decreasing (Fig. \ref{fig:Corot3x}).

The evolution of the system was simulated considering four different initial periods for the stellar rotation: 1, 2, 4, and 8 days. A broad range of initial values was considered so as to put into evidence the strong convergence of the star's rotation period to an equilibrium value when braking is taken into account. The adopted value for the relaxation factor ($\gamma = 55 s^{-1}$) and the other initial values were chosen so that the system reproduces the currently observed values after about 6 Gyr. The empirical formulas given in \SFM \ allow us to convert, for a given system, the relaxation factor to the usual quality factor $Q$ or its modified counterpart $Q'$; the adopted value of $\gamma$ corresponds, for this star, to Q= $6 \times 10^6$.
An extended analysis of the survival of short-period exoplanets allowed Hansen (2012) to estimate that the dissipation values $Q$ of solar-type host stars lie in the interval $8\times 10^5 - 4\times 10^7$ 
(that is, $ 0.6 \times 10^7 < Q' < 30  \times 10^7$); see Hansen (2012; fig. 16). 

{For comparison, the results of simulations with the same initial conditions, but without the braking, are also shown. They are very instructive}: In the absence of braking, the rotation of the star tends to the super-synchronous stationary solution of Darwin theories given by Eq. (\ref{eq:offset}). 
Besides, a decrease of the orbital period is not {visible}; there is a quick variation in the first few Gyr of the simulation, when the rotation of the star is still adjusting itself to the orbital period of the companion, but once this phase is over, the orbital period, that is, the semi-major axis, remains almost constant. The comparison to the complete simulation puts into evidence the role played by the stellar wind braking on the inward migration of the companion. The rightmost panel of fig. \ref{fig:Corot3x} shows that the same effect is also observed in the circularization of the orbit, which becomes faster in the presence of braking,  {as a consequence of the continuous loss of angular momentum by the system}.

Finally, we have to mention that in the above calculations and in those reported in the forthcoming sections, the contribution of the tide raised by the star on the companion {did not contribute to the results.} 
The rotational evolution of brown dwarfs is not well known (see Reiners and Basri, 2008) and we assumed 
that the rotation of the companion becomes stationary on a very short time scale (a few Myr) and that, very soon in the system's evolution, the tide in the companion stops making a significant contribution.


\section{Kepler-75, CoRoT-27 and their giant hot Jupiters}\label{Sec:C27K75}

\begin{figure}[t]
\centerline{\hbox{
\includegraphics[width=6cm,clip=]{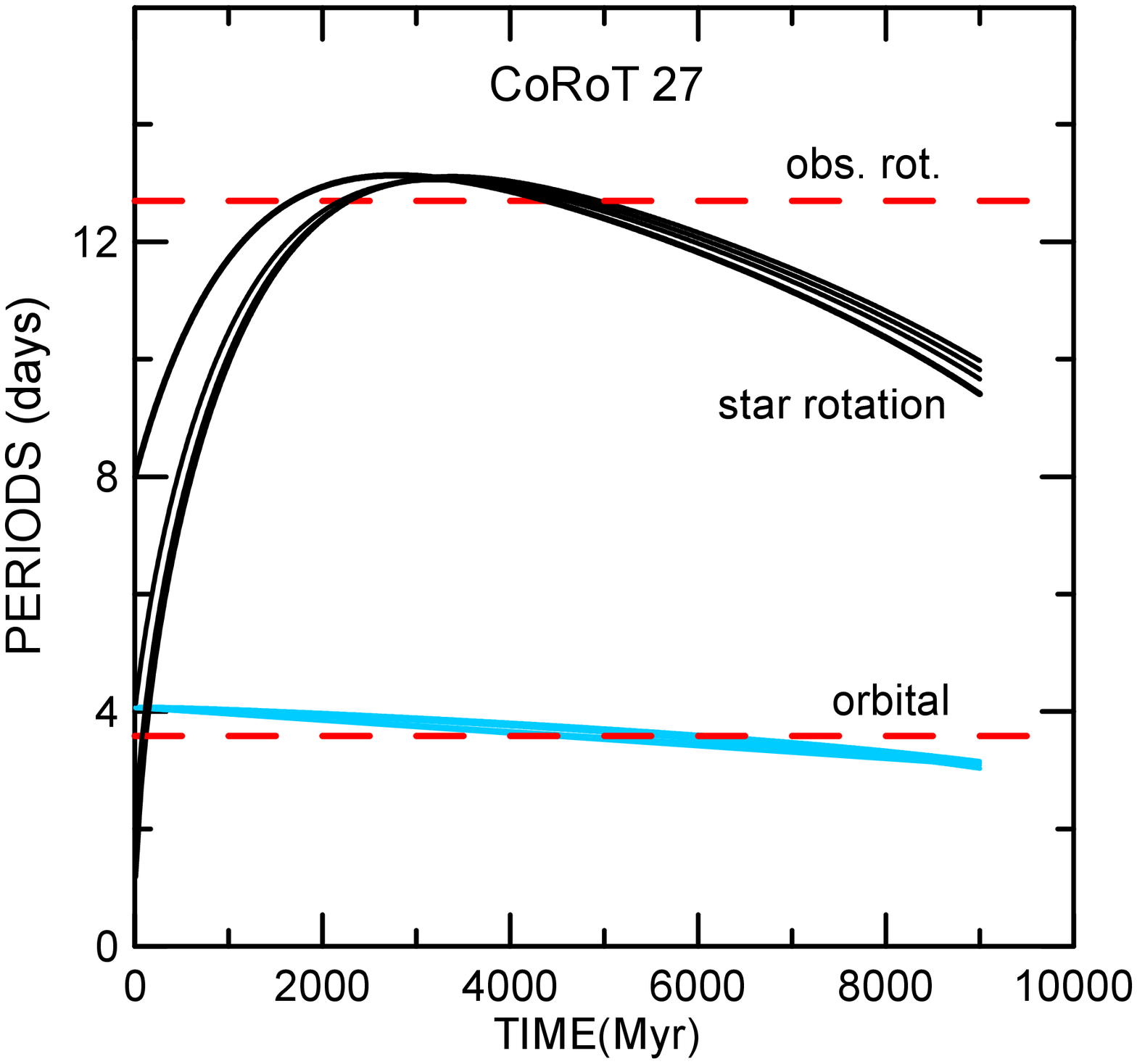}\hspace{5mm}
\includegraphics[width=6cm,clip=]{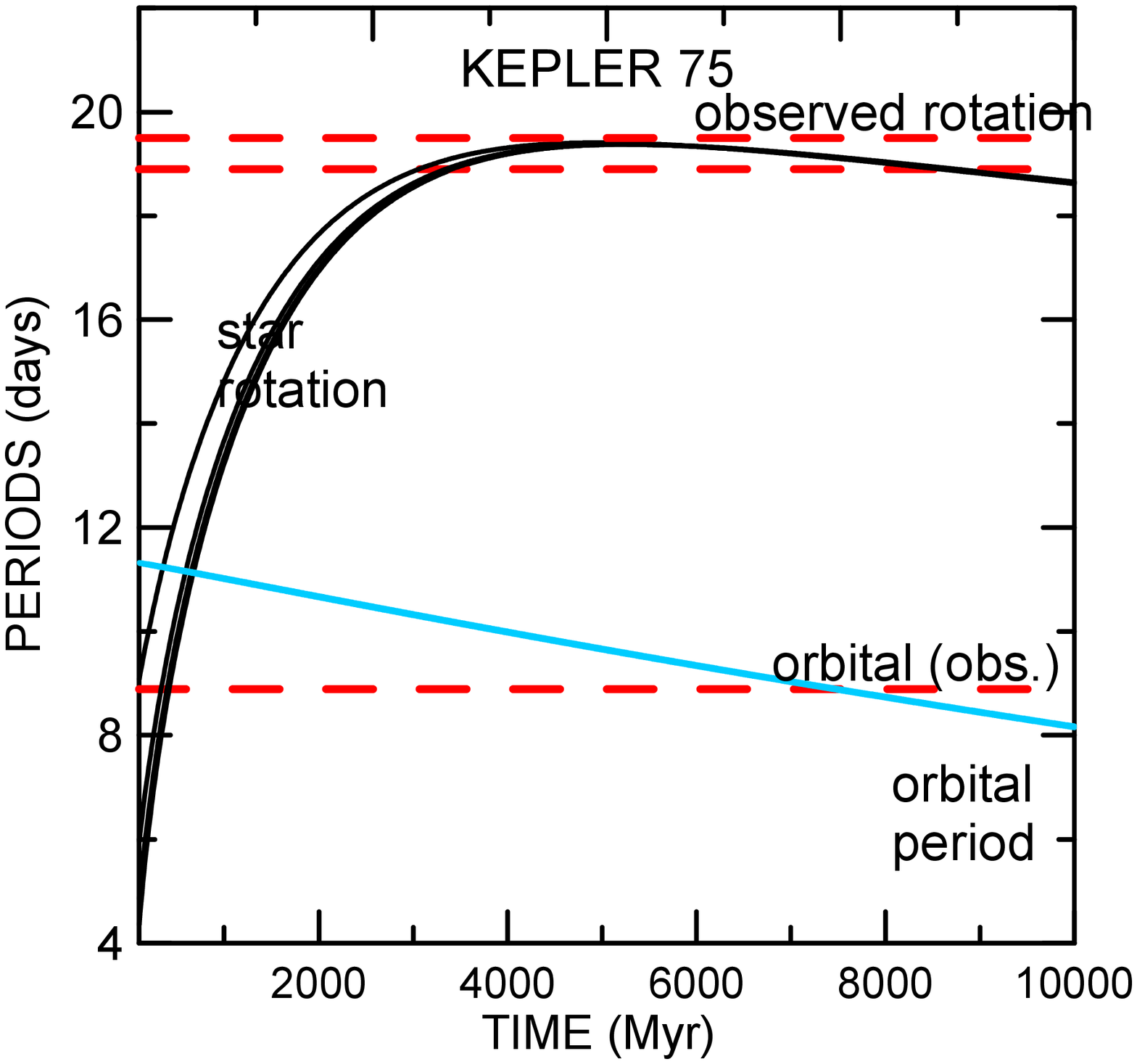}
}}
\caption{Evolution of the rotation of the star and the orbital period of the planet under the joint action of the tidal torque on the star and the stellar wind braking. {Left:} CoRoT-27; {\it Right:} Kepler-75. Initial periods of stellar rotation are 1, 2, 4, and 8 days. The red dashed lines show the observed values of the orbital period of the planet and the photometric period of the star. }
\label{fig:K75C27}
\end{figure}

CoRoT-27 and Kepler-75 are two stars with transiting hot super-Jupiters with masses close to $10 M_{\rm Jup}$. 
CoRoT-27b evolves in an almost circular orbit at $a = 0.0476 \pm 0.0066$ AU and behaves like the above discussed paradigm, but with a less steep stationarization timescale.
Kepler-75b evolves in a higher orbit ($a = 0.080 \pm 0.005$ AU) but, because of its high eccentricity ($e=0.57\pm 0.1$), it plunges to very close to the star at every pericenter passage. 
Evolution of the orbital period of the planets and the rotation period of these stars under the joint action of the tidal torque on the stars and the stellar wind braking is shown in fig. \ref{fig:K75C27}. 

The relaxation values adopted in these cases were, respectively, $\gamma=90$ and $\gamma=108$ s$^{-1}$. These values, as well as the initial orbital elements of both simulations, come from the solution of the inverse problem. 
They indicate dissipations smaller than in the CoRoT candidate LRc06E21637; however,  
much larger or much smaller values lead to evolutions that are either too fast or too slow. 
Smaller values of $\gamma$ mean higher tidal dissipation; tides dominate over the braking and the stars should have in the past an excessively slow rotation, which is not compatible with the standard scenarios admitted for star formation and evolution. 
Larger values of $\gamma$ mean lower tidal dissipation; in such cases, the evolution of the stellar rotation is dominated by the braking and the star's rotation should reach the current rotation value in a time shorter than the published age of the star. 

Other host stars in this class are HAT-P-20 and HAT-P-21,{ the latter being in a significantly non-circular orbit: $e = 0.228 \pm 0.016$.} Their evolution is similar to those already shown. However, in these cases the error in the determined rotation period is large and the fitting of the evolution to it is less informative. The corresponding plots are omitted. We only mention that the evolution of the rotation period obtained for HAT-P-20 with a more complex model treating the convective and radiative parts of the star separately (Penev et al. 2012, Fig. 1), after the contraction of the star is finished, is almost equal to that obtained with the model adopted in this paper. 

{We may also mention the star KOI-205. Notwithstanding the large-mass companion, a brown dwarf, the semi-major axis is large ($a=0.0997\pm 0.0013$ AU) and the tidal torques are almost ineffective. Unlike the others, the system is still evolving towards a stationary solution. We obtain, for this K0 star, $\gamma > 100$. For smaller values (larger dissipations), the star is unable to reach the observed slow rotation in a time less than the determined age limit of 8 Gyr. It is worth adding that an age smaller than $6-7$ Gyr is impossible to explain with the usual braking models.}

\section{The short-living CoRoT-2 and CoRoT-18}\label{Sec:C2C18}

\begin{figure}[t]
\centerline{\hbox{
\includegraphics[width=5cm,clip=]{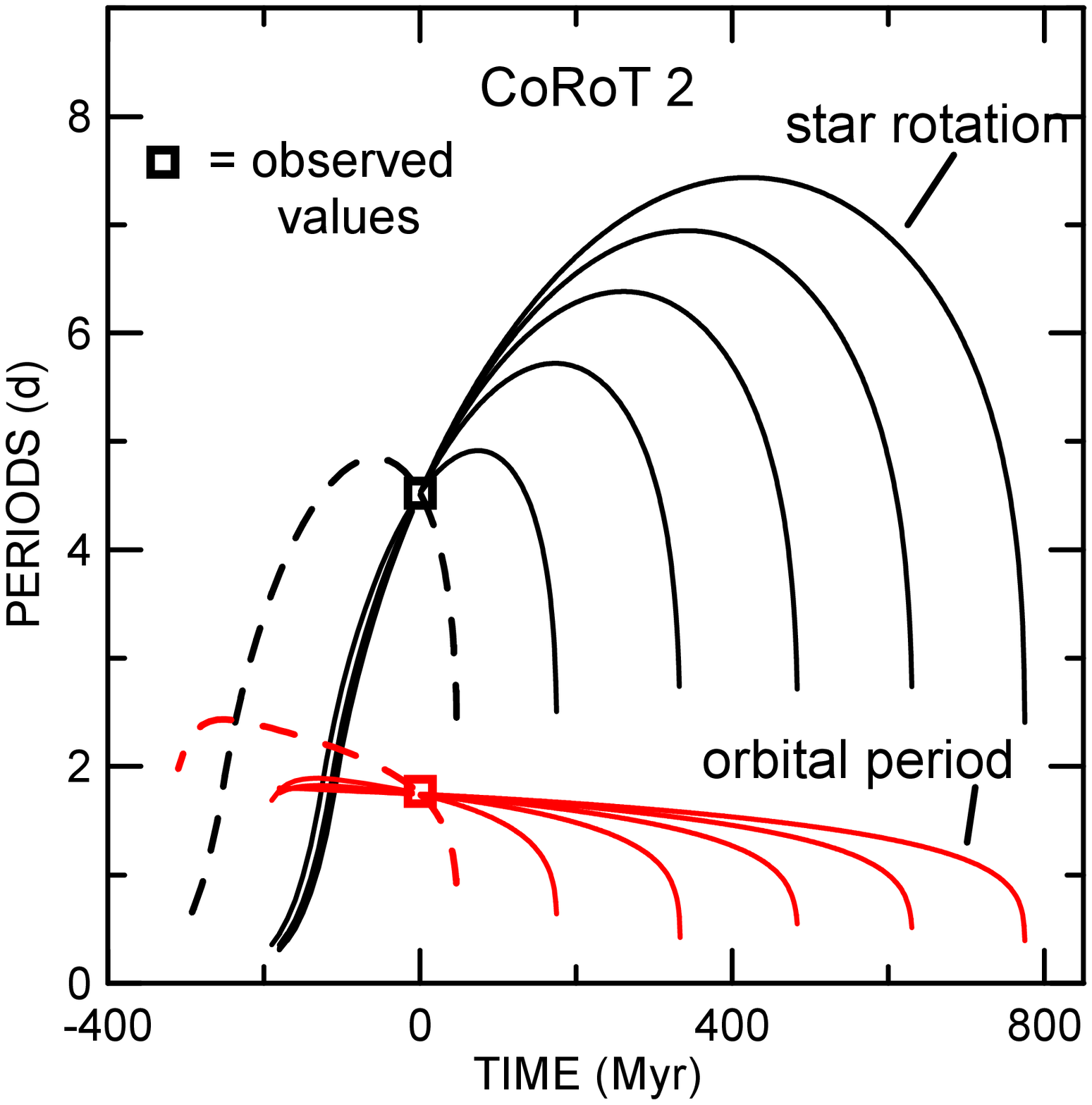}  \hspace{7mm}
\includegraphics[width=5.1cm,clip=]{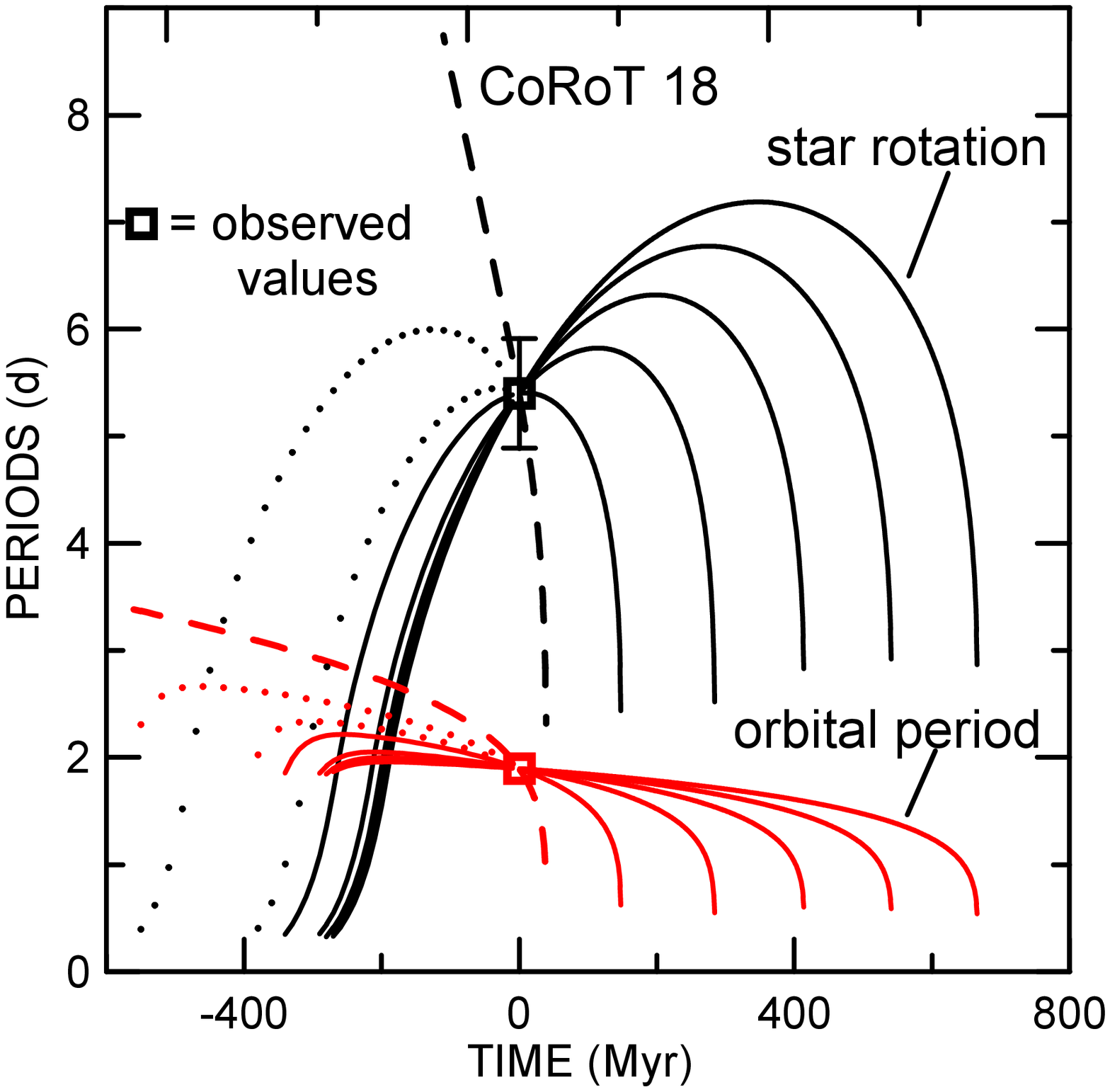}}}
\caption{Evolution of the star's rotation period (black) and of the orbital period of the planet (red) under the joint action of the tidal torque on the star and the stellar wind braking. 
{\it Left:} CoRoT-2; {\it Right:} CoRoT-18.  
Adopted relaxation factors: 5 (dashed line); 10, 15 (dots); 20, 40, 60, 80, 100 (solid lines) s$^{-1}$.
Squares: Initial conditions (observed).}
\label{fig:C2C18}
\end{figure}

These two systems are very young and have not had enough time to evolve to a stationary solution. The study of their evolution may then follow a different approach. We start from the present conditions and propagate them forward and backward. The results are shown in the two panels of fig. \ref{fig:C2C18} where the present conditions are shown by a square. Given the impossibility of solving for the tidal relaxation factor, we adopted a wide range of values covering the values found for the other solar-type stars: $\gamma = 20, 40, 60, 80, 100 $s$^{-1}$ (solid lines in fig. \ref{fig:C2C18}). Just for the sake of comparison, we also added results with $\gamma = 5 $s$^{-1}$ (dashed line). In all evolution paths shown, the initial (current) eccentricity $e=0.01$ was adopted.

Fig. \ref{fig:C2C18} {\it left} shows that the {evolutionary tracks of CoRoT-2 for $\gamma > 20 {\rm s}^-1$ converge, in the past, to very fast rotations, in about 0.2 Gyr.} Its X-ray luminosity and strong chromospheric emission indicate a young stellar age of about 200-300 Myr (Poppenhaeger and Wolk, 2014). 
The planet CoRoT-2b will fall into the star in less than 1 Gyr (0.2 Gyr if $\gamma=20 $s$^{-1}$ and 0.8 Gyr if $\gamma=100 $s$^{-1}$). The age estimate is very robust. For example, if the initial (current) eccentricity is taken as $e=0.06$ instead of $e=0.01$, the age grows by no more than some 10-20 \%. The expected lifetime, however, decreases in this case by about  50 \% { because of the increase of the dissipation with the eccentricity}. More influential is the current period, but the strong activity of the star allowed for a very precise determination of the photometric period. We also do not know the primordial stellar rotation, which adds a non-statistical error in the age determination. 

Fig. \ref{fig:C2C18} {\it right} shows that the age of CoRoT-18 is about 0.3-0.4 Gyr. Again, the same sources of inaccuracy discussed above exist. We notice that the photometric period has not been determined in this case with the same relative precision as for CoRoT-2. The characteristic feature in this case is the steep evolution of the rotation period and the impending fall of the planet into the star in the solution obtained with $\gamma=5 $s$^{-1}$ (dashed line). 
The comparison of the solutions with $\gamma=5 $s$^{-1}$ and $\gamma=10 $s$^{-1}$ (given by the leftmost dotted line) indicates that for some values in this interval, the evolutionary path of the rotation may become compatible with ages much larger than 0.4 Gyr. 
However, these values correspond to extremely high tidal dissipation, not expected for stars; 
They correspond, for this star, to Q= $0.4-0.8 \times 10^6$. The expected range after Hansen's (2012) work is 
$ 0.8 \times 10^6 < Q < 40 \times 10^6$. Thus, solutions showing an age larger than 0.5 Gyr for CoRoT-18 are not actually expected.

The ages of CoRoT-2 and CoRoT-18 determined by isochrone fitting (Brown, 2014) are much larger than those guessed above. Brown's larger ages are not precluded by the results obtained for the evolution of these planetary systems, but would indicate dissipation values larger than those of the stars in the sample studied by Hansen (2012). The ages of these stars determined by gyrochronology are $0.2 \pm 0.1$ and $0.25 \pm 0.3$ Gyr, respectively (Brown, 2014).

{Finally, it is worth mentioning that the planets CoRoT-2b and CoRoT-18b are somewhat bloated (radii resp. $1.465\pm0.029$ and  $1.31 \pm 0.18$ R$_{\rm Jup}$), unlike the other hot Jupiters considered in this study whose radii do not differ by more than 10 percent from Jupiter's radius.}

\section{The F stars CoRoT-15, KELT-1 and HAT-P-2}\label{Sec:F-stars}

F stars with masses larger than 1.3 $M_\odot$ are not expected to have a wind braking (see the distribution of the present rotations of the NGC 6811 stars with B-V$<0.55$ in Meibom et al. 2011). Therefore, their rotation in the presence of large close-in planets must be that predicted by Darwinian theories. Their rotations must be synchronous with the orbital motions for very small eccentricities and super-synchronous, i.e. faster  than synchronous, for increasing eccentricities (as in the comparative simulations without braking of fig. \ref{fig:Corot3x} {\it left}). 
Two examples are shown in table 1: CoRoT-15 and KELT-1 whose companions are close-in brown dwarfs in nearly circular orbits. In both cases the observed values of the rotation period are consistent with synchronization. The rotation of KELT-1 comes from the determined $v\sin I$; the observed Rossiter-McLaughlin effect for this star indicates a projected spin-orbit alignment angle of 2 $\pm $16 deg, consistent with a zero obliquity for KELT-1. The $v.\sin I$ for CoRoT-15 indicates a projected rotation slower than the orbital period. No observations allowed the estimation of $I$ and so an actual period closer to the orbital period is possible. It is worth mentioning that the autocorrelation study of the light curve showed two peaks very close to the orbital period of CoRoT-15b . 

The third F star in table 1 is HAT-P-2, which has been classed as  marginally tidal-stable by Levrard (2009). This star is in the transition zone between stars with and without stellar wind braking. Several authors have noted that, following the Darwinian model, because of the high eccentricity ($e=0.517 \pm 0.003$), this star should have a rotation period roughly equal to half the orbital period of the companion. The highest observed value was considered as an indication that the stationarization of the star's rotation is not yet reached. This opinion is consistent with the fact that the simulations of the evolution of this system show that the stationarization of the star's rotation is indeed very slow. An alternative hypothesis is that the star is the seat of some braking, even a small one (e.g. $f_P=0.15$), and becomes stationary with a period larger than the supersynchronous Darwinian limit, but smaller than the synchronous value.

\section{The puzzling case CoRoT-14}\label{Sec:C14}

\begin{figure}[t]
\centerline{\hbox{
\includegraphics[width=6cm,clip=]{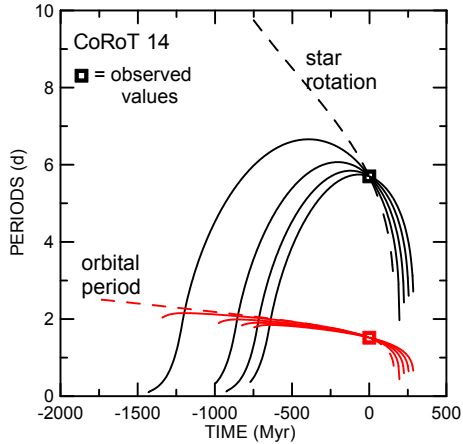}}}
\caption{Evolution of the rotation of CoRoT-14 (black) and of the orbital period of the planet (red) under the joint action of the tidal torque on the star and the star's wind braking as constrained by the current rotation period of the star: 5.6 $\pm$ 0.5 d.
Adopted relaxation factors (in s$^{-1}$): 250 (dashed line), 300, 350, 400, and 450 (solid lines). 
Squares: Initial conditions (the observed values).}
\label{fig:C14}
\end{figure}

Another example is the star CoRoT-14 classed as F9V with large reddening and $B-V = 0.85 \pm 0.22$.  The rotation period of this star is several times larger than the orbital period of the companion, leading us to believe that wind braking may take place. 
Since the age of the star is largely unknown, its evolution is studied using the same approach as in Sec. \ref{Sec:C2C18}. 
The results for the typical values of $\gamma$ used in the study of the other systems show the same behavior: The transfer of angular momentum from the orbit to the star via tidal torque is so large that the star's rotation is being continuously accelerated (as shown by the dashed line in Fig. \ref{fig:C14}). The magnitude of the acceleration is such that, at some time in the past, the rotation of the star should be retrograde. This is an unlikely physical situation that could, however, result as consequence of disk migration in binary systems whose orbital plane is uncorrelated with the spin axes of the individual stars (Batygin, 
2012). We have to find alternatives. One of them is to assume a large $\gamma$. 
In the results shown in fig. \ref{fig:C14} for $\gamma$ = 250, 300, 350, 400,  and 450 s$^{-1}$, we see that values of $\gamma$ equal to $300$ s$^{-1}$ or larger (solid lines) allow us to have a system that is evolving smoothly from an ordinary situation in the distant past in which the star has a fast rotation up to reach the present condition.  $\gamma$ = 300 s$^{-1}$ is roughly equivalent, for this system,  to $Q = 4 \times 10^6$, a value that is still in the range of values determined by Hansen (2012) for solar-type stars, but larger than the maximum $10^6$ adopted by Dobbs-Dixon et al (2004). 
{It is worth noting that the dissipation models studied by Barker and Ogilvie (2009) show large variations in the dissipation of F stars, thus allowing for larger values of $\gamma$.}

The analysis does not allow us to set an upper limit for the age of the star.

We cannot discard the possibility of having some inaccuracy in the data used.
A faster current rotation would imply in smaller values of $\gamma$; the same is true for a larger semi-major axis. We have reanalyzed the photometric data of this star obtained with CoRoT and the radial velocity measurements. We obtained (and adopted) a different mass: $6.95 \pm 0.5 M_{\rm Jup}$ (instead of the $7.6 \pm 0.6 M_{\rm Jup}$ given by Tingley et al. 2011).  A self-correlation analysis allowed us also to obtain some alternative determinations of  the rotation period smaller than the $5.6 \pm 0.6$ d obtained from Fourier analysis of the data (or 5.7 days {\it cf.} Tingley et al. 2011), but the precision of the data does not allow us to choose one of them. 

Another possibility to consider is that the measured surface rotation does not reflect the rotation of the inner radiative zone, which could be much faster. The convective zone of this F9V star is small and maybe the timescale of the angular momentum exchanges between the two zones is too large to allow a larger reduction of the primordial rotation of the whole star. A mere reduction of the rotation period by 20-30 percent would be enough to give for this star an evolution pattern similar to CoRoT-2 and CoRoT-18 and with the same relaxation factors (i.e. dissipation) adopted in those cases. 

Otherwise, we have to look for some drastic changes in the scenario. We may, for instance, assume that the system is young and that the primordial period of the star was rather large and is decreasing because of the strong tidal torque. 
We may also assume that the current system is the result of a late migration of the planet to the close-in neighborhood of the star, reaching it when, due to wind braking, the star was already rotating much more slowly than its primordial rotation. 
Some prosaic hypotheses are yet possible. The star possesses significant excess emission in the mid-infrared (see Krivov et al. 2011). We may assume that the star is young and the seat of  wind interactions with the inner disk that are able to remove part of the angular momentum in a more efficient way. 

We have also conducted some numerical experiments without including the star's wind braking. The results are very similar but, in such a case, not even values as high as $\gamma=1000 s^{-1}$ are capable of providing an acceptable solution. 

\section{Limits of the tidal interaction}\label{Sec:geral}

\begin{figure}[t]
\centerline{\hbox{
\includegraphics[width=6cm,clip=]{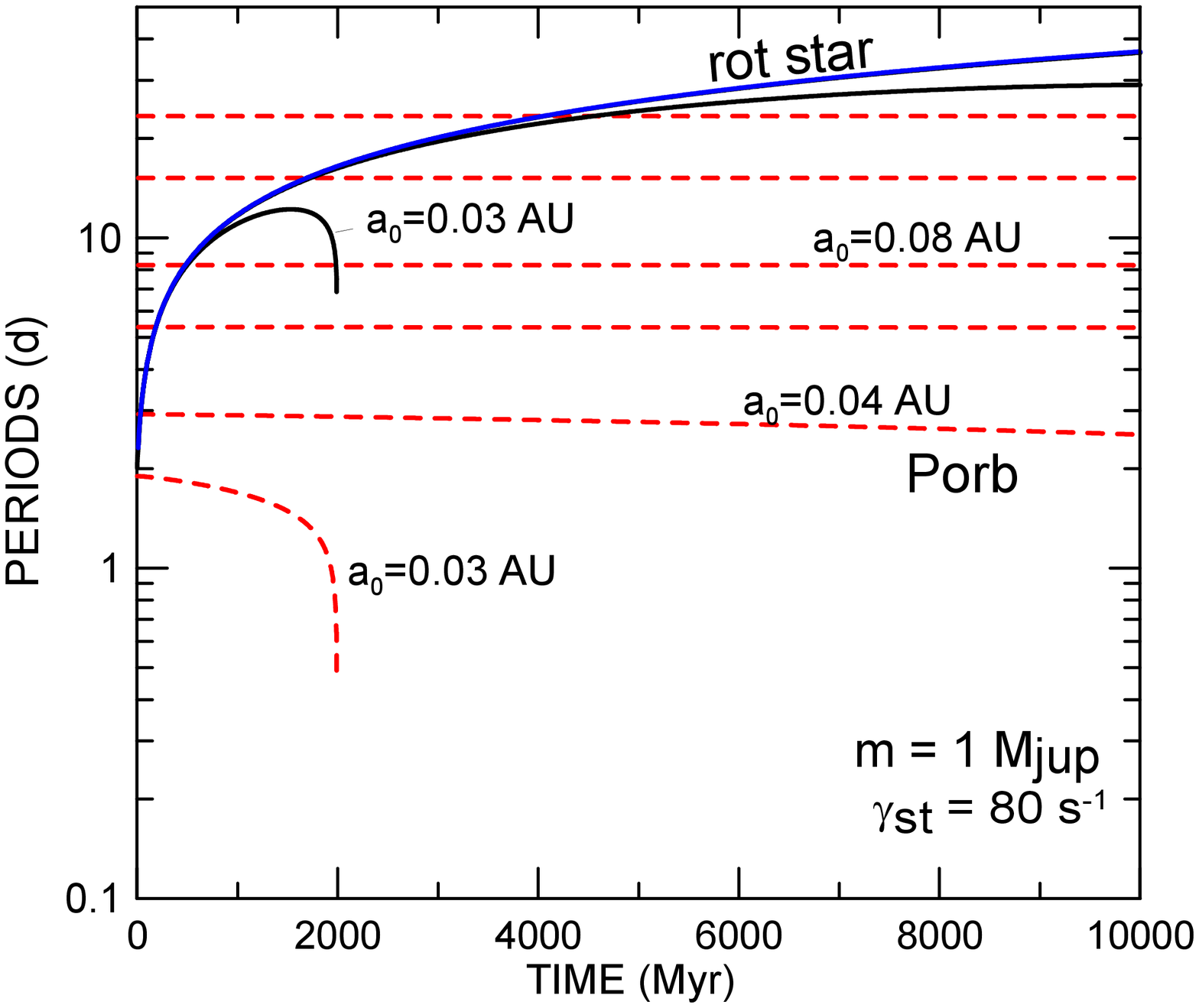}  \hspace{7mm}
\includegraphics[width=6cm,clip=]{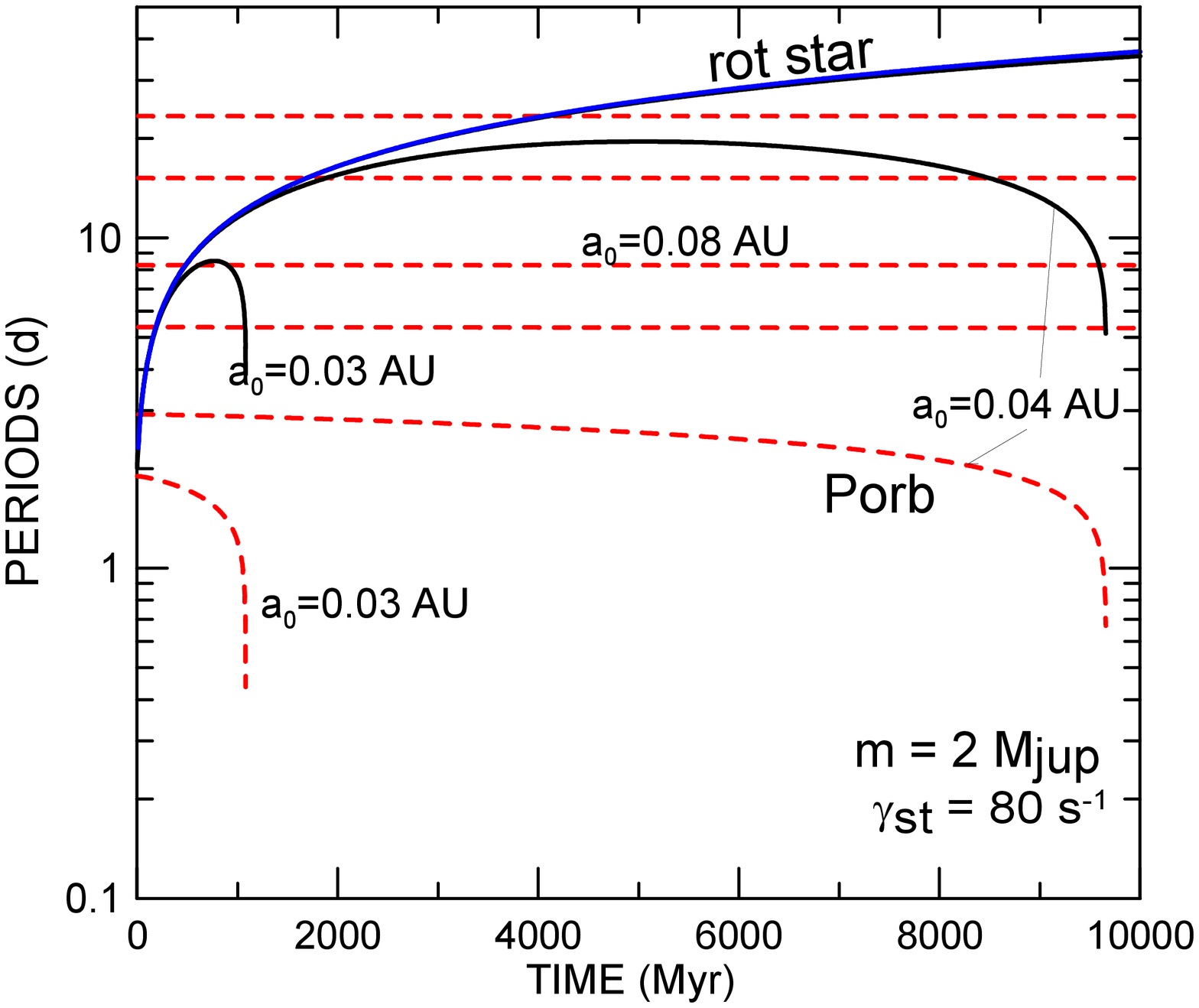}}}
\centerline{\hbox{
\includegraphics[width=6cm,clip=]{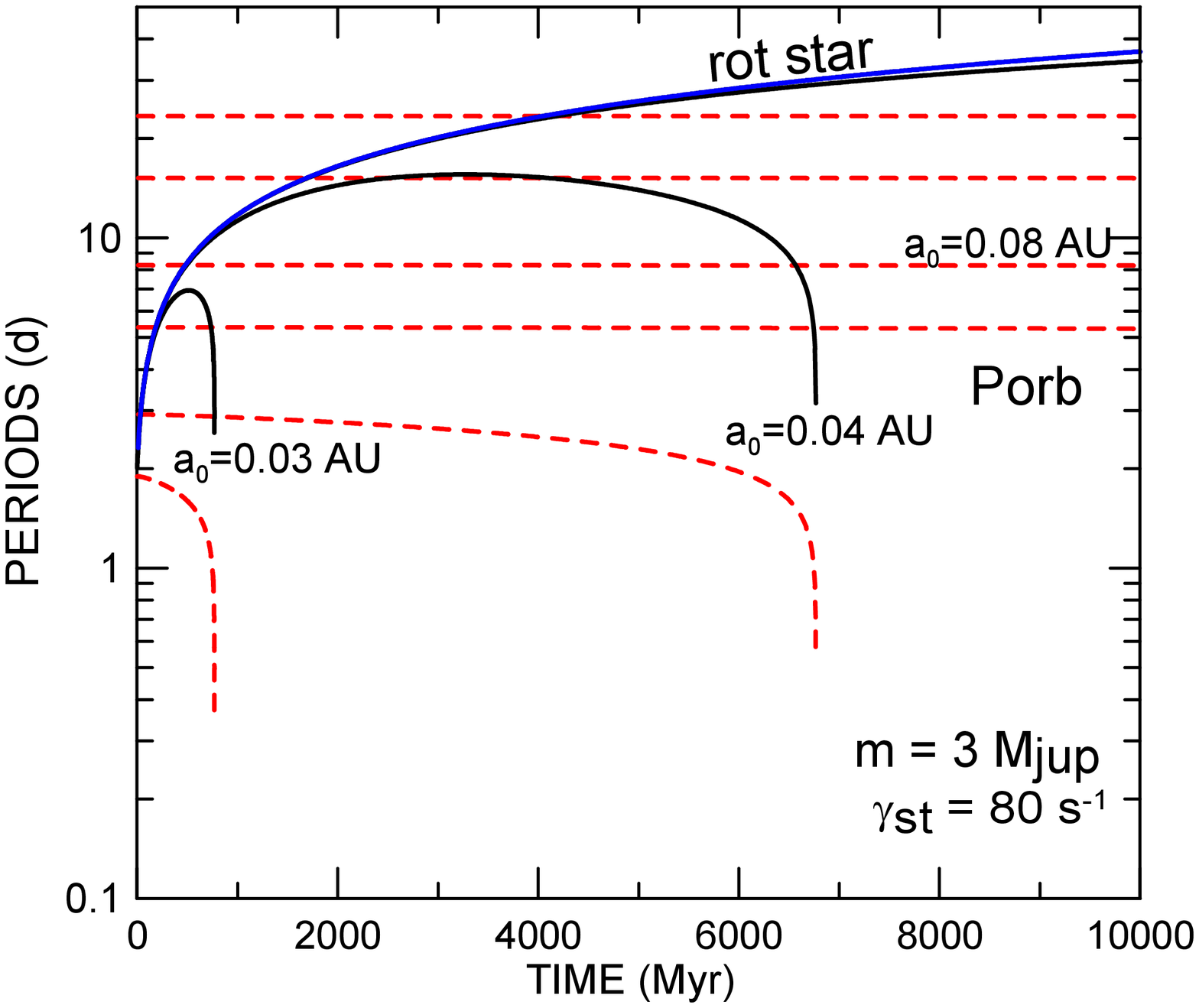}  \hspace{7mm}
\includegraphics[width=6cm,clip=]{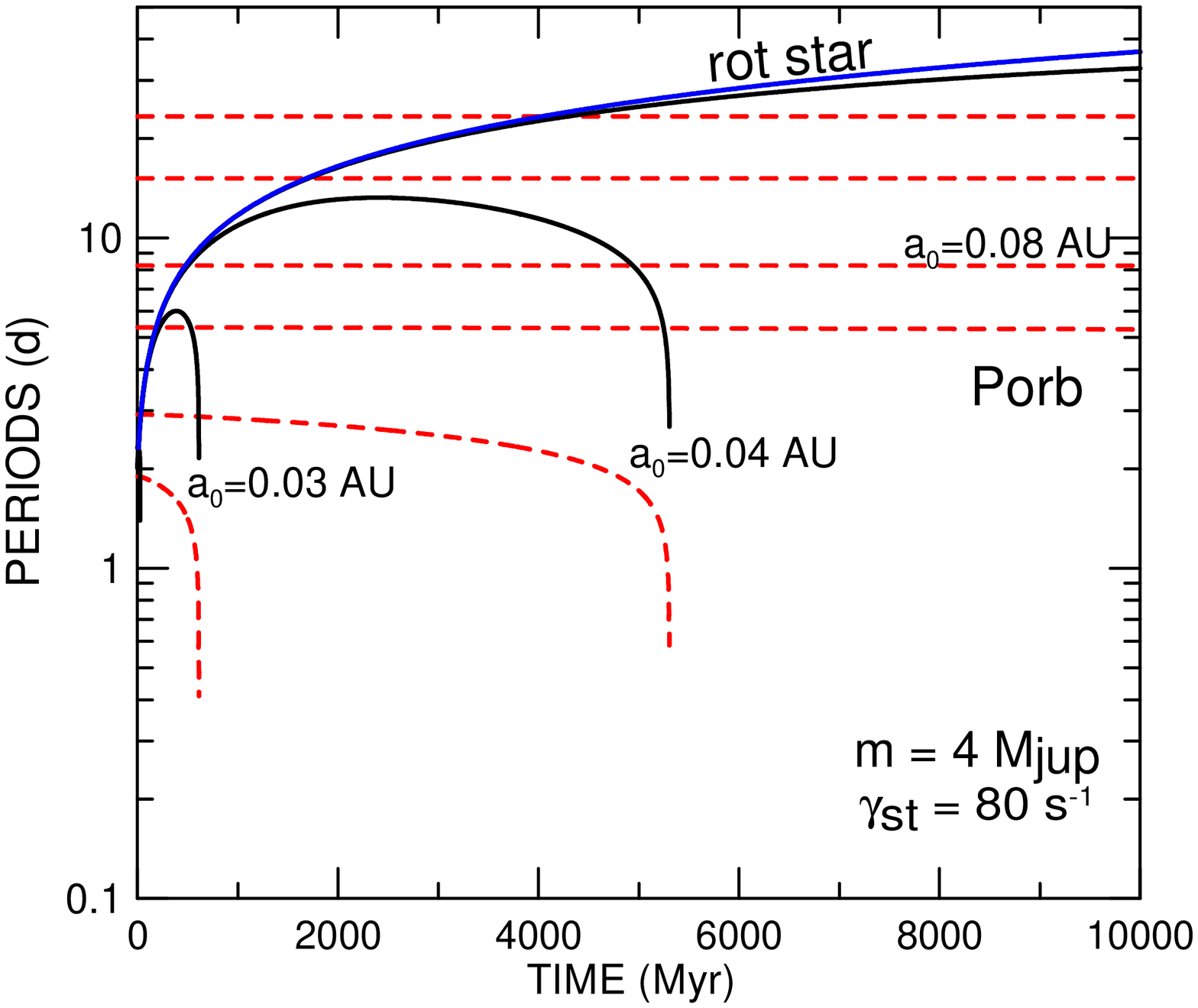}}}
\caption{Evolution of the rotation of a 1 $M_\odot$ star (black/solid line) and of the orbital period of companion hot Jupiters (red/dashed line) with masses resp. 1, 2, 3, 
and 4 $M_{\rm Jup}$
The solutions shown correspond to initial circular orbits with semi-axes 0.03, 0.04, 0.06, 0.08, 0.12, and 0.16 AU (Initial periods between 1.9 and 23.3 days). The blue envelope corresponds to a negligible tide. Adopted relaxation factor: 80 s$^{-1}$. Adopted braking: $f_P$=1.}
\label{fig:quadra}
\end{figure}

It is important to know the location of the boundary after which the tidal interaction is no longer able to transfer angular momentum from the orbit of the companion to the rotation of the star. For that sake, we consider systems with one star like the Sun in the center and one hot Jupiter in orbit around it. The results are shown in fig. \ref{fig:quadra} whose panels show the evolution of the rotation period of the star and the orbital period of the companion for several initial values of the semi-major axis. In all examples, we consider that the star's initial rotation period is 2 days and the studied planets have masses between 1 and 4 Jupiter mass and radius equal to 1 Jupiter radius; some simulations were also done assuming a bloated planet, with radius up to 1.6 Jupiter radius, but no appreciable change was seen in the  results. The initial orbits were circular and the dissipation in the planet did not play any role in the evolution of the considered periods. The chosen initial period of 2 days is near the median of rotation periods observed in clusters of stars in the early main sequence (see Gallet and Bouvier, 2013); as seen in the examples studied in Secs. \ref{Sec:C3x} and \ref{Sec:C27K75}, the fast rise of the star's rotation period in the first half Gyr is such that the evolution curves quickly converge to neighboring values. 
 
In all cases of very close planets, the observed pattern is the same: The orbital period decreases, indicating an inward spiraling of the planet up to its fall into the star. The rotation period of the star initially is increasing due to the star's wind braking, but at some point the braking is overcome by the tidal transfer of angular momentum from the planet's orbit and the rotation period starts decreasing, up to the eventual fall of the planet into the star when a rapid decrease in the rotation period is seen.  

In the case of more distant planets (initial distances larger than 0.06 AU), the variation of the orbital period is very small and the stellar rotation is driven almost only by the stellar wind braking. 

\begin{figure}[t]
\centerline{\hbox{
\includegraphics[width=8cm,clip=]{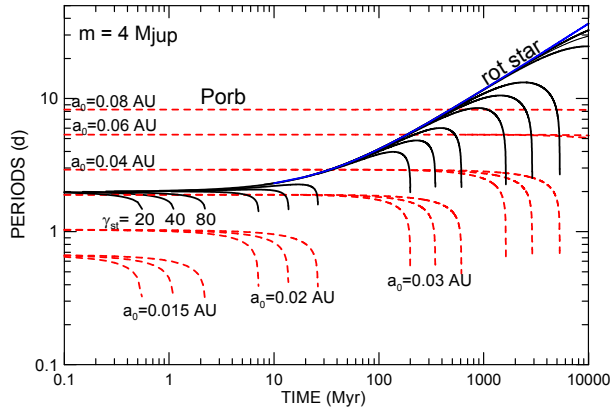}}}
\caption{Evolution of the rotation of a 1 $M_\odot$ star (black/solid line) and of the orbital period of a companion hot Jupiter (red/dashed line) with mass 4 $M_{\rm jup}$. The solutions shown correspond to initial circular orbits with semi-axes between 0.015 and 0.08 AU (Initial periods between 0.66 and 8.25 days). Adopted relaxation factors: 20, 40, 80 s$^{-1}$. The blue envelope corresponds to negligible tide. The black lines are not labeled, but they are easily identified by comparison to the labeled red lines with the same final time. }
\label{fig:geral}
\end{figure}

In the cases shown in fig. \ref{fig:quadra}, the adopted relaxation factor is $\gamma =80$ s$^{-1}$. For a solar-type star with a companion in a 5-day orbit, this relaxation factor roughly corresponds to $Q=5-6$ million (we recall that a general relation linking $Q$ and $\gamma$ does not exist). We have also done simulations with stronger dissipations: $\gamma =$ 20 and 40 s$^{-1}$. The results are shown in fig. \ref{fig:geral} where we adopted a logarithmic timescale to better allow the comparison of the results for the three considered relaxation factors. The logarithmic scale also allows us to see the case of closer planets, with lifetimes of a few Myr. {These results may be compared to the similar ones obtained by Bolmont et al. (2012) for planets with 1 $M_{\rm Jup}$ around a 1-$M_\odot$ star. There, the tidal action is visible even for planets initially at distances as great as 0.06 AU because the dissipation adopted was 1000 times larger than Hansen's (2010) determinations.}

We have also considered some alternative initial scenarios. 
First was the case where initial orbits are non-circular. For $e=0.2$, the result is qualitatively the same as shown in figs. \ref{fig:quadra} and \ref{fig:geral}, but with a faster timescale and lifetimes about 40-50 percent shorter than in the circular case. For instance, the rightmost falling termination shown in fig. \ref{fig:geral} is, in the eccentric orbit case, at 3.6 Gyr instead of 5.3 Gyr.
Next, we considered a star with a much slower initial rotation period (8 days). Again, the results repeat what is shown in figure \ref{fig:geral} with the obvious difference that the rotation periods of the star are initially larger. However, after the first Gyr, they almost coincide with the results obtained with a faster initial rotation. 

One important feature shown by the above figures (and also by figs. \ref{fig:C2C18} and \ref{fig:C14}) is the huge increase in the stellar rotation speed when the planet falls into the star. As suggested by Teitler and K\"onigl (2014), this fact can explain the lack of close-in Kepler objects of interest around stars with short rotation periods. 
The star's rotation is in some sense reset to a primordial-like rotation when a close-in hot Jupiter is tidally driven to fall into it.

\section{Conclusions}

The consideration of tidal torques is not enough to understand the current rotation of solar-type stars hosting big close-in companions. It is also necessary to take into account the fact that, in these stars, the stellar wind is continuously consuming its angular momentum. Thus, while the final state of a system with a close-in companion affected only by tidal torques is a star rotating faster than, or at least synchronized with, the orbital motion of the companion, the presence of a stellar wind braking displaces the final state towards much longer stellar rotation periods.  When the braking drives the stellar rotation to rotation periods larger than the orbital period of the close-in companion, the tidal torques will transfer angular momentum from the orbit to the rotation of the star, accelerating it. The total torque vanishes at periods larger than the orbital period.

In this paper, we have gathered the main formulas used to compute these two effects and applied them to study some selected planetary systems. 
{We initially considered one large sample including all known systems with one close-in companion 
with mass larger than 3 $M_{\rm Jup}$ around a main sequence star. 
However, most of the systems in this first selection had to be discarded because for many of them the basic data on the stars are unknown. 
In several cases not even the mass of the central star is known.
One critical datum for the analyses done here is the rotation of the star. 
We privileged systems for which the rotation period of the star was determined from photometric studies and is thus free of the $\sin I$ indetermination. 
But, in some cases, we have to use spectroscopic determinations of the rotation speed and assume that $I=90^\circ$. They are identified in Table I. 
We also discarded a few systems for which the observation of the Rossiter-McLaughlin effect indicated a large projected obliquity or for which the spectral class and age of the central star were not known.}

The main studied cases were some systems with a central G star and age larger than some Gyr. In these cases we could reconstruct the possible past evolution of the orbital elements and the rotation of the star. We have shown that, independently of the initial rotation of the star, these  solutions converge to a situation of equilibrium between the opposite tendencies imposed on the star's rotation  by the magnetic wind braking and by the tidal torques due to the companion.
The sample also included some systems with a central F star and ages larger than 1 Gyr. It is symptomatic that, in these cases, the rotation period does not differ significantly from the orbital period of the companion. This has been interpreted as resulting from the absence of braking. In the absence of braking, the classical tide theories predict that the rotation period of the star evolves toward a stationary rotation slightly faster than the orbital motion.   
Finally, we have considered some younger systems and discussed the possibility of determining their ages from the simulations of the past evolution. In two cases, CoRoT-2 and CoRoT-18, the evolutionary paths through the current data converge to a fast rotation at some time in the past, thus giving an indication of the age of the star. 

As an important by-product, we have determined the tidal relaxation factor $\gamma$ 
of some of these stars. The values obtained lie in the interval $\gamma = 25-110 $s$^{-1}$ with a possible outlier, CoRoT-14, for which the photometric period given in the discovery paper leads to $\gamma > 250$ s$^{-1}$. However, the assumption of a stellar rotation period 20-30 percent smaller than the observed surface rotation period gives
results for this system that are consistent with the value of the relaxation factor for the other stars studied here.

In general, the values obtained are in agreement with the limit $\gamma > 30$ s$^{-1}$ for solar-type stars proposed in Ferraz-Mello (2013)  and the corresponding estimates of the quality factor $Q$ ($8 \times 10^5 < Q < 4 \times 10^7$) obtained by Hansen (2012) from an extended analysis of the survival  of short-period planets. We recall that the previous results of Hansen (2010), obtained without taking the braking into account, were smaller (corresponding to $\gamma = 5-25$ s$^{-1}$). These smaller values can be understood. As shown in fig. \ref{fig:Corot3x}, the change in the stellar rotation due to the stellar wind braking accelerates the decrease of the orbital period -- that is, it shortens the lifetime of the planets. Therefore, the reproduction of the rates of circularization and fall of the planets without consideration of the stellar wind braking implies in dissipation values larger than the actual ones (i.e. smaller values of $\gamma$). 

As far as the accuracy of the adopted $\gamma$ is concerned, we remember that the tidal torque is affected by the errors in the determination of the star's mass. 
The adopted $\gamma$ also depends on the generally ill-determined age and period of rotation of the stars. One more source of inaccuracy is the braking always assumed with the strength given by Bouvier's formula, which is also just an estimation that is not necessarily followed closely by every star considered. 
In addition, the past scenarios of the various cases shown in this paper are ideal extrapolations that can deviate from reality due to changes impossible to detect (e.g. the past presence of another companion in the same system).
Given these caveats, the individual values adopted for $\gamma$ are only rough estimates. 
We may expect an improvement of these results with the measurements to be made by \textit {PLATO} (Rauer, 2014). On one hand, the strong focus of \textit {PLATO } on bright targets may allow the determination of stellar ages with high accuracy (10 percent). On the other hand, the frequency analysis of the photometric data can also provide information about the interior rotation of some stars and thus open the way to the use of more complete models taking into account the rotation mixing inside the stars.

One consequence of the above results is that the presence of a significant tidal torque does not allow us to use gyrochronology rules (Barnes, 2007) to estimate the age of a star hosting a large close-in companion (see Brown, 2014). Simulations using hypothetical values show that a companion with mass over 1 $M_{\rm Jup}$ in an orbit within 0.04 AU of a solar-type star (i.e. with period less than 3 days) produces  an evolution in the rotation of the star different from that predicted by the Skumanich's law (see figs. \ref{fig:quadra} and \ref{fig:geral}). In these cases, at variance with the good results obtained for single stars (see do Nascimento et al. 2014), gyrochronology rules would systematically indicate agers for old stars with massive companions that are smaller than the actual ones. This difference may be small for stars with Jupiter-like companions near the given distance limit, but for closer planets (or planets with much larger masses), the observed pattern is very different from the standard evolutions shown by the results for negligible tide, and the error in the determined age may be large. 
{In contrast, for solar-type stars with planets not so large or at larger distances, the gyrochronology rules can be used for the age determination with the same confidence that they are used for single stars.}

\acknowledgements
The CoRoT space mission has been developed and operated by CNES, with the contribution of Austria, Belgium, Brazil, ESA, Germany, and Spain. 
This investigation was supported by grants CNPq 306146/2010-0, FAPESP 2011/52096-6 and by INCT Inespa\c co procs. FAPESP 2008/57866-1 and CNPq 574004/2008-4. 
The help of Mr. E.S. Pereira in the calculation of some complementary examples is acknowledged.

\end{document}